\documentclass[useAMS]{mn2e}
\usepackage{graphicx,amssymb,amsmath}
\usepackage{hyperref,xcolor}
\usepackage{subfigure}
\usepackage{pdflscape}
\usepackage{mathrsfs}

\newcommand{\scs}{\scriptsize}

\title[Open clusters Stock 2, NGC 2168, 6475, 6991 and 7762] {Comprehensive abundance analysis of red giants in the open clusters Stock 2, NGC 2168, 6475, 6991 and 7762 }
\author[A. B. S. Reddy and D. L. Lambert]
  { Arumalla B. S. Reddy$^{1}$\thanks{E-mail: bala@astro.as.utexas.edu},
  and David L. Lambert$^1$ \\
   $^1$W.J. McDonald Observatory and Department of Astronomy, The University of Texas at Austin, Austin, TX 78712, USA \\ }

\begin{document}

 \date{Accepted 2019 February 07. Received 2019 February 06; in original form 2019 January 07}

\pagerange{\pageref{firstpage}--\pageref{lastpage}} \pubyear{2016}

\maketitle

\label{firstpage}

\begin{abstract}
We have analysed high-dispersion echelle spectra ($R = 60000$) of red giant members of five open clusters to derive abundances for many
elements from Na to Eu. The [Fe/H] values are $-0.06\pm0.03$ for Stock 2, $-0.11\pm0.03$ for NGC 2168, $-0.01\pm0.03$ for NGC 6475,
$0.00\pm0.03$ for NGC 6991 and $-0.07\pm0.03$ for NGC 7662. Sodium is enriched in the giants relative to the abundance expected of main
sequence stars of the same metallicity. This enrichment of [Na/Fe] by about $+0.25$ attributed to the first dredge-up  is discussed in the light
of theoretical predictions and recently published  abundance determinations. Abundance ratios [El/Fe] for other elements are with very few exceptions
equal to those of field giants and dwarfs, i.e., [El/Fe] $\simeq 0.00$ for [Fe/H] $\sim 0.0$. An exception is the  overabundance of La, Ce, Nd and Sm in NGC 6991 but
this is consistent with our previous demonstration that the abundances of these $s$-process products vary by about $\pm0.2$ among clusters of the
same [Fe/H], a variation found also among field giants and dwarfs.

\end{abstract}

\begin{keywords}
Galaxy: abundances -- Galaxy: disc -- Galaxy: kinematics and dynamics -- (Galaxy:) open clusters and associations: general -- stars: abundances
 -- open clusters: individual: Stock 2, NGC 2168, NGC 6475, NGC 6991 and NGC 7762
\end{keywords}

\section{Introduction} 

Observations of stars in Galactic open clusters  yield information on two distinct areas of astrophysics: the evolution of stars and the structure of the Galactic disc. In this paper, the fifth in a series on the chemical compositions of red giants in clusters, chemical compositions are provided for giants in five clusters bringing the total number of clusters studied to 33.  Compositions of giants across this sample are applied to two investigations -- one concerning  stellar evolution and a second involving the origin of field stars in the Galactic disc.

In the area of stellar evolution, the focus here is on sodium abundances in giants. Sodium abundance of a giant is predicted to be enhanced over the initial value thanks to sodium brought to the surface by the first dredge-up experienced as the star evolves up the red giant branch. This paper examines if the observed increase in the sodium abundance agrees with that predicted for post first dredge-up giants. 

Concerning field stars a key premise is that open clusters  dissolve over time with their stars joining the population of field stars in the Galactic disc. Then, one may ask if it is possible to assign  field stars to their now-dissolved open clusters. This exercise would be aided greatly if each open cluster had a distinctive chemical compositions, i.e., one or more chemical tags (Freeman \& Bland-Hawthorn 2002). In an earlier paper (Lambert \& Reddy 2016) based on the compositions of giants in open clusters, it was observed that the abundance of heavy elements (e.g., La and Ce) whose synthesis is attributed to the main $s$-process in AGB stars may have different abundances in clusters with otherwise identical compositions. This observation about abundances of heavy elements is pursued further in this paper as part of the dream of establishing a practical chemical tag.

This is the first paper in our series on open clusters to be completed following the second Gaia data release (Gaia DR2, Gaia Collaboration et
al. 2018) which consists of celestial positions, parallaxes, broad-band photometry for sources brighter than G-band magnitude 21. Gaia DR2 also
provides radial velocities with typical errors of 0.3$-$1.8 km s$^{-1}$ for stars brighter than G$_{RVS}=$ 12 mag, and high-quality transverse velocities at a precision of 0.07 mas yr$^{-1}$ (G$<$15 mag) to 3 mas yr$^{-1}$ (G$<$21) for an unprecedentedly large number of stars. With Gaia DR2,  recent studies have presented revised cluster distances, proper motions, radial velocities, and investigated the distribution of open clusters in the 6D phase space (Cantat-Gaudin et al. 2018; Soubiran et al. 2018). None of the five clusters in this paper, with the exception of NGC 7762, was considered previously with Gaia DR2. The photometric and astrometric parameters of cluster stars collected from Gaia DR2 allow us to confirm the cluster membership of red giants analysed in this paper.

This paper is organized as follows: In Section 2 we describe observations, data reduction and radial velocity measurements. Section 3 is devoted to the abundance analysis and Section 4 to discussing revised cluster parameters. We present in Section 5 comparison with abundance determinations in the literature. In Section 6 we discuss the Na abundances in light of predictions for Na enrichment resulting from the first dredge-up.  Section 7 discusses the heavy elements and the possible chemical tag they provide. Section 8 provided concluding remarks.

\begin{table*}
\centering
\caption{The journal of the observations for the cluster members.} 
\label{log_observations}
\begin{tabular}{lccccccccclc}   \hline
\multicolumn{1}{l}{Cluster}& \multicolumn{1}{c}{Star} & \multicolumn{1}{c}{$\alpha(2000.0)$}& \multicolumn{1}{c}{$\delta(2000.0)$}&  
\multicolumn{1}{c}{V}& \multicolumn{1}{c}{B-V} & \multicolumn{1}{c}{V-K$_{\rm s}$} &  \multicolumn{1}{c}{J-K$_{\rm s}$} &
\multicolumn{1}{c}{$RV_{\rm helio}$} & \multicolumn{1}{l}{S/N at} & \multicolumn{1}{l}{Date of} & \multicolumn{1}{c}{Exp. time}  \\
\multicolumn{1}{c}{}& \multicolumn{1}{c}{}& \multicolumn{1}{c}{(hh mm ss)}& \multicolumn{1}{c}{($\degr$ $\arcmin$ $\arcsec$)}&
\multicolumn{1}{c}{(mag)}& \multicolumn{1}{c}{ } & \multicolumn{1}{c}{ }& \multicolumn{1}{c}{ } & 
\multicolumn{1}{c}{(km s$^{-1}$)} & \multicolumn{1}{l}{6000 \AA } & \multicolumn{1}{l}{observation} & \multicolumn{1}{c}{(sec)}  \\
\hline
  Stock 2  & 43 & 02 13 28.79 & $+$59 11 45.59 &  7.58 &$+$1.24 &$+$2.89 &$+$0.68 &$+$08.8$\pm$0.1 & 350 & 17-10-2016 & 1$\times$1200  \\
     & 1011 & 02 11 11.63 & $+$59 58 51.25 &  8.17 &$+$1.24 &$+$2.99 &$+$0.70 &$+$08.6$\pm$0.1 & 380 & 17-10-2016 & 2$\times$1200   \\
     & 1082 & 02 15 16.71 & $+$59 20 05.82 &  8.45 &$+$1.33 &$+$3.40 &$+$0.74 &$+$04.4$\pm$0.1 & 240 & 17-10-2016 & 2$\times$1800   \\
NGC 2168 &  81 & 06 09 00.86 & $+$24 15 56.14 &  8.57 &$+$1.38 &$+$3.34 &$+$0.80 &$-$07.6$\pm$0.1 & 210 & 18-10-2016 & 1$\times$1500 \\
          & 310 & 06 09 15.93 & $+$24 25 40.38 &  7.42 &$+$1.13 &$+$2.55 &$+$0.62 &$-$07.5$\pm$0.1 & 260 & 18-10-2016 & 1$\times$960 \\
          & 662 & 06 08 50.78 & $+$24 30 02.48 &  8.53 &$+$1.27 &$+$3.29 &$+$0.83 &$-$08.3$\pm$0.1 & 235 & 18-10-2016 & 1$\times$1800 \\
 NGC 6475 &HD~162587 & 17 53 23.47 & $-$34 53 42.44 &  5.60 &$+$1.09 &$+$2.57 &$+$0.73 &$-$13.4$\pm$0.1 & 280 & 18-10-2016 & 1$\times$240 \\
   & HD~162587$^{1}$ & 17 53 23.47 & $-$34 53 42.44 & $\ldots$ & $\ldots$ & $\ldots$ & $\ldots$ &$-$18.5$\pm$0.2 & 480 & 14-11-2016 & 1$\times$600 \\
   & HD~162587$^{2}$ & 17 53 23.47 & $-$34 53 42.44 & $\ldots$ & $\ldots$ & $\ldots$ & $\ldots$ &$-$03.0$\pm$0.2 & 480 & 14-11-2016 & 1$\times$600 \\
    & HD~162391 & 17 52 19.76 & $-$34 25 00.64 &  5.85 &$+$1.10 &$+$2.41 &$+$0.77 &$-$14.5$\pm$0.1 & 275 & 18-10-2016 & 1$\times$300 \\
    & HD~162496 & 17 52 49.22 & $-$34 06 53.40 &  6.06 &$+$1.24 &$+$2.72 &$+$0.82 &$-$14.3$\pm$0.1 & 400 & 19-10-2016 & 1$\times$600 \\ 
 NGC 6991 &  22 & 20 53 16.77 & $+$47 27 14.58 & 11.16 &$+$0.92 &$+$2.52 &$+$0.60 &$-$12.9$\pm$0.1 & 150 & 18-10-2016 & 2$\times$1800 \\
          &  67 & 20 54 29.82 & $+$47 28 03.18 & 09.42 &$+$1.04 &$+$2.51 &$+$0.62 &$-$12.8$\pm$0.1 & 160 & 16-09-2016 & 1$\times$1800 \\
          & 100 & 20 55 03.96 & $+$47 19 20.13 & 09.91 &$+$1.06 &$+$2.29 &$+$0.59 &$-$12.2$\pm$0.1 & 190 & 15-09-2016 & 2$\times$1200 \\
          & 131 & 20 55 42.70 & $+$47 22 32.70 & 09.67 &$+$1.05 &$+$2.24 &$+$0.55 &$-$12.4$\pm$0.1 & 170 & 15-09-2016 & 1$\times$1800 \\
 NGC 7762 &  35 & 23 49 21.69 & $+$68 01 01.97 & 11.66 &$+$1.71 &$+$4.54 &$+$0.94 &$-$49.2$\pm$0.1 & 135 & 20-10-2016 & 3$\times$1200 \\
          &  91 & 23 50 31.42 & $+$68 01 41.51 & 11.73 &$+$1.75 &$+$4.21 &$+$0.91 &$-$47.7$\pm$0.1 & 135 & 20-10-2016 & 3$\times$1800 \\
          & 110 & 23 49 06.13 & $+$67 59 08.58 & 12.56 &$+$1.71 &$+$4.14 &$+$0.87 &$-$45.7$\pm$0.1 & 110 & 19-10-2016 & 4$\times$1800 \\
\hline
\end{tabular} 
\flushleft{ Note: 1 \& 2 are the individual components of the observed spectrum of HD\,162587.}
\end{table*}

\section{Observations and data reduction}

Selection of red giants for subsequent spectroscopic observations was made by identifying stars having common proper motions at the putative 
red giant clump in the cluster colour-magnitude diagram (CMD). The CMDs created from the available BV and Gaia DR2 photometric magnitudes along with  
our program stars marked by red squares are shown in Figure \ref{cmdcluster} and Figure \ref{cmd6991} whose details will be presented in  later sections. 

High-resolution and high signal-to-noise (S/N) ratio optical spectra of 18 giant stars in six OCs, namely Stock 2 (3 giants), NGC 2168 (3 giants), NGC 6475 (3 giants), NGC 6991 (4 giants) and NGC 7762 (3 giants) were observed in 2016 September, October and November with the Robert G. Tull coud\'{e} cross-dispersed echelle spectrograph (Tull et al. 1995) at the 2.7-m Harlan J. Smith telescope of the McDonald observatory. On all occasions we employed the camera with a 2048$\times$2048 24 $\mu$m pixel CCD detector and 52.67 grooves mm$^{-1}$ echelle grating with exposures centred at 5060 \AA\ in order 69. We secured two to four exposures of cluster giants in our target list (Table \ref{log_observations}) with each exposure limited to 20-30 min to minimize the influence of cosmic rays and to acquire a global S/N of above 100.

Each night's observing routine included five zero second exposures (bias frames), 15 quartz lamp exposures (flat frames), and 2-3 exposures of Th-Ar lamp spectra along with observing the targets. The two-dimensional spectral frames were extracted to one-dimensional images in multiple steps using various routines available within the \textit{imred} and \textit{echelle} packages of the standard spectral reduction software {\small IRAF}\footnote{IRAF is a general purpose software system for the reduction and analysis of astronomical data distributed by NOAO, which is operated by the Association of Universities for Research in Astronomy, Inc. under cooperative agreement with the National Science Foundation.}. In short, each exposure of the program star was de-trended by removing bias level and the scattered light, and then divided by the normalized flat field. The individual echelle orders were traced, extracted to one-dimensional spectral format and then wavelength calibrated using Th-Ar lamp spectra as a reference. Observed spectra cover the wavelength range 3600$-$9800 \AA\ over multiple echelle orders in a single exposure, with gaps between echelle orders of 10-120 \AA\ for wavelengths longer than 6100 \AA, which is sufficient to perform an abundance analysis of elements sampling all the major processes of stellar nucleosynthesis.

All the spectra correspond to a resolving power of $R$ $=$ 60,000 (5 km s$^{-1}$) as measured by the FWHM of Th {\scs I} lines in comparison spectra. The spectrographic setup was stable throughout the night as inferred by the lack of significant systematic shift of Th {\scs I} lines in comparison spectra ($\lesssim$\,1 m\AA) taken at the beginning and end of the night. Our wavelength scale based on the thorium-argon spectra is accurate within a root-mean-square scatter of 3 m\AA. Multiple spectra of a star were combined to acquire a single spectra whose S/N ratios greatly exceed 100 per pixel, permitting reliable estimate of line equivalent widths (EWs) down to the 3 m\AA\ level. The combined spectra of each star has S/N values over 100 across many echelle orders, but for wavelengths shorter than about 4000 \AA\ the S/N ratio gradually drops and reaches a value of about 15 around 3600 \AA\ region.

The spectum of each red giant was trimmed, normalized interactively to unity.  The radial velocity (RV) was measured from a set of 20 lines with well defined line cores. The observed RVs were transformed to the heliocentric velocities using the {\it rvcorrect} routine in {\scs IRAF}.
The methods of observations, data reduction and RV measurements are described in depth in Reddy, Giridhar \& Lambert (2012, 2013, 2015). The properties of the cluster giants observed in this study are summarized in Table \ref{log_observations} together with the available optical and 2MASS\footnote{\url{http://irsa.ipac.caltech.edu/applications/Gator}} photometry (Cutri et al. 2003)\footnote{Originally published by the University of Massachusetts and Infrared Processing and Analysis Center (IPAC)/ California Institute of Technology.}, computed heliocentric RVs, and S/N ratios measured around 6000 \AA.

\subsection{Radial velocities and cluster membership} \label{rv_mem}

Extensive studies of proper motions and radial velocities exist in the literature for the members of OCs NGC 2168, NGC 6475, NGC 6991 and NGC 7762 excluding Stock 2. Although the proper motion data is available for all stars in the field of Stock 2, only the giant stars have been subjected to RV measurements in the literature. The agreement of measured RVs between ours and the literature sources is excellent given the slightly larger measurement errors in the literature. We summarize below the comparison of our results with those from literature to verify cluster membership of each stellar target.

\subsubsection{NGC 2168}
The mean RV of the cluster NGC 2168 was determined previously using a sample of dwarf and red giant members in three different studies: Barrado y Navascu\'{e}s, Deliyannis \& Stauffer (2001, hereafter BN01), Geller et al. (2010) and Mermilliod, Mayor \& Udry (2008). BN01 measured radial velocities to an accuracy of $\sigma=$1 km s$^{-1}$ from  high-resolution spectra (R$\sim$20,000) of 39 main-sequence dwarfs showing no sign of binarity, obtaining a mean RV of the cluster of $-$8.0$\pm$1.5 km s$^{-1}$. 
NGC 2168 has been studied extensively as part of the WIYN Open Cluster Survey (WOCS; Mathieu 2000); consequently, membership has been established for the main-sequence dwarfs via both proper motions (McNamara \& Sekiguchi 1986a) and radial velocities (Geller et al. 2010).  
Geller et al. (2010) determined the cluster mean RV of $-$8.16$\pm$0.05 km s$^{-1}$ using the spectra of 344 solar-type stars within the magnitude range 13.0 $<$\,V\,$<$ 16.5, whose individual RVs were estimated to a precision of $\pm$0.5 km s$^{-1}$. 
Since we selected only giants for abundance analysis, we have no stars in common with the above two studies for a direct star-to-star comparison of velocities. But a comparison of the cluster mean RVs obtained between the studies is  useful. The mean RV of $-$7.8$\pm$0.3 km s$^{-1}$ ($\sigma=$0.1 km s$^{-1}$) obtained for NGC 2168 using the three red giants in our study is compatible with the mean values computed previously by BN01 and Geller et al. (2010) from the cluster dwarfs. 

The study of Mermilliod et al. (2008) involving all the red giants in common with our work has resulted a cluster mean RV of $-$8.4$\pm$0.5 km s$^{-1}$ ($\sigma=$1.1 km s$^{-1}$) which suggest a fair agreement between the analyses. Had we compared our RV estimates of individual red giants with those measured by Mermilliod et al., the fair agreement would be essentially unchanged given the measurement uncertainties between the studies. Excellent agreement of the cluster mean RV between the red giants and main-sequence dwarfs of NGC 2168 strengthens identification of  the giant stars selected for abundance analysis in this paper as cluster members.

\subsubsection{NGC 6475}
Mermilliod, Mayor \& Udry (2009) reported radial velocities from  CORAVEL spectrovelocimeter observations of 75 main-sequence stars in the field of the open cluster NGC 6475. Excluding 9 spectroscopic binaries, the mean RV of $-$14.8$\pm$0.2 km s$^{-1}$ was computed using 33 potential members of NGC 6475. With a typical uncertainty of 0.5$-$2.5 km s$^{-1}$ in individual measurements, the standard deviation in the mean RV of the above sample is about $\sigma=$1.3 km s$^{-1}$. In a different study, Mermilliod et al. (2008) reported radial velocities of two of the red giants in this cluster among which the star HD 162587 was classified as a double-lined spectroscopic binary. 

Two separate observations of the star HD 162587 were made, as the spectrum acquired on October 18, 2016 exhibits a clear sign of core splitting of the spectral lines that prompted us to reobserve the star on November 14, 2016 whose spectrum shows the evidence of double-lined nature of HD 162587. 
Excluding the star HD 162587 in table \ref{log_observations}, we estimated the cluster mean RV of $-$14.4$\pm$0.1 km s$^{-1}$ ($\sigma=$0.1 km s$^{-1}$) using two red giants. An excellent agreement of the mean RVs computed independently from the red giants and a large sample of main-sequence dwarfs (Mermilliod et al. 2009) show that the giant stars included in our study are members of NGC 6475.

\subsubsection{NGC 6991 and NGC 7762}
We measured a mean radial velocity of $-$12.6$\pm$0.3 km s$^{-1}$ (4 giants) and $-$47.5$\pm$1.4 km s$^{-1}$ (3 giants) for NGC 6991 and NGC 7762, respectively. The velocity dispersion found among these cluster giants is typical of what is seen among members of most Galactic OCs. The typical velocity dispersion of a virialized OC is of the order of 1 km s$^{-1}$ or less with the supplementary condition that the presence of undetected binaries can inflate the measured velocity dispersion of a cluster by many km s$^{-1}$ (Girard et al. 1989; Geller, Latham \& Mathieu 2015). 
Although the intrinsic velocity dispersion of NGC 7762 is slightly higher than the typical of OCs, such a large radial velocity dispersion was measured previously among stars in this cluster (Casamiquela et al. 2016; Carraro, Semenko, \& Villanova 2016). 
Therefore, we consider that the giant stars observed in this study are potential members of the respective OCs in Table \ref{log_observations}.

The radial velocities for some of our cluster members were measured previously in two different studies: Casamiquela et al. (2016) have reported RVs for 6 stars each in NGC 6991 and NGC 7762 from the high-resolution spectra (R$\geq$62,000) acquired at the Spanish Observatories. For stars in common between the studies, the RV measurements are in fair agreement with mean differences of $+$0.2$\pm$0.2 km s$^{-1}$ (3 stars) and $+$0.2 km s$^{-1}$ (1 star) for stars in NGC 6991 and NGC 7762, respectively. Carraro et al. (2016) have measured RVs and metallicities using the medium resolution spectra (R$\geq$13,000) of 8 giants in NGC 7762, of which we have two stars in common. Although our RV estimates agree well for one star ($\#$110), the other star ($\#$91) in NGC 7762 exceeds Carraro et al.'s value by $+$2.5 km s$^{-1}$. However, noting the larger velocity differences in the range $-$2.4 to $+$2.6 km s$^{-1}$ for five stars in common between the analyses of Carraro et al. and Casamiquela et al. (see Table 2 in Carraro et al. 2016), the larger velocity difference found here for one star is not surprising.

\subsubsection{Stock 2}
Although the radial velocity information exists only for a few giants in the field of Stock 2, accurate proper motions of the cluster members within an area of 70$\arcmin\times\,$70$\arcmin$ were measured previously by Spagna et al. (2009). Using a sample of 275 main-sequence dwarfs satisfying the conditions of i) measurement errors within 1.55 mas yr$^{-1}$ each in proper motions and ii) the membership probabilities P$_{\rm r}\geq$\,90\%, Spagna et al. measured the cluster mean proper motions of $\mu_{\alpha}$\,cos $\delta=+$16.27$\pm$0.09 mas yr$^{-1}$ and $\mu_{\delta}=-$13.33$\pm$0.07 mas yr$^{-1}$. 
 
The selection of three red giants for spectroscopic observations was made on the basis of close agreement of their proper motions with that of the cluster mean. We computed the cluster mean proper motions of $\mu_{\alpha}$\,cos $\delta=+$16.55$\pm$0.62 mas yr$^{-1}$ and $\mu_{\delta}=-$13.25$\pm$0.31 mas yr$^{-1}$ from the three red giants having a typical proper motions uncertainty of 0.08 mas yr$^{-1}$ each from the Gaia astrometry (Gaia Collaboration et al. 2018).
These values compare well with the values obtained from the main-sequence dwarfs in the literature. Although two of the cluster giants with IDs 43 and 1082 were classified as spectroscopic binaries by Mermilliod et al. (2008), we found no evidence of double-lined spectrum. Hence, we consider both the stars as single-lined binaries and conclude that the spectral energy distribution of the secondary companion must have negligible effect on the abundance analysis of these stars. Two of the cluster giants in Table \ref{log_observations} were measured to have consistent RVs while the RV of one star 1082 differs by 4 km s$^{-1}$ from the cluster mean RV of $+$8.5$\pm$0.8 km s$^{-1}$ ($\sigma=$2.9 km s$^{-1}$) obtained from Mermilliod et al.'s measurements. However, on the account of binarity and RV difference of 3.4 km s$^{-1}$ between ours and Mermilliod et al.'s value, we regard the star 1082 as a potential cluster member.

\begin{table*}
\centering
\begin{minipage}{150mm}
\caption{Photometric and spectroscopic atmospheric parameters for open cluster members analysed in this study. }
\label{stellar_param}
\begin{tabular}{lcccccccccc}  \hline
\multicolumn{1}{l}{Cluster}& \multicolumn{1}{c}{Star ID} & \multicolumn{3}{c}{T$^{\rm phot}_{\rm eff}$ (K)} &
\multicolumn{1}{c}{$\log g^{(B-V)}_{\rm phot}$}& \multicolumn{1}{c}{T$^{\rm spec}_{\rm eff}$}& 
\multicolumn{1}{c}{$\log g^{\rm spec}$}& \multicolumn{1}{c}{$\xi^{\rm spec}_{t}$}&
\multicolumn{2}{c}{$\log(L/L_\odot)$} \\  
\cline{3-5}
\cline{10-11}
\multicolumn{1}{c}{}& \multicolumn{1}{c}{} & \multicolumn{1}{c}{(B-V)}& (V-K)& (J-K)& 
\multicolumn{1}{c}{(cm s$^{-2}$)}& \multicolumn{1}{c}{(K)} & \multicolumn{1}{c}{(cm s$^{-2}$)}& 
\multicolumn{1}{c}{(km s$^{-1}$)}& \multicolumn{1}{c}{phot} & \multicolumn{1}{c}{spec} \\
\hline

  Stock 2 & 43 & 5067 & 5311 & 4916 & 2.27 & 4925 & 2.00 & 1.75 & 2.51 & 2.72  \\
      & 1011 & 5067 & 5184 & 4953 & 2.46 & 4900 & 2.30 & 1.57 & 2.31 & 2.43  \\
      & 1082 & 4938 & 4696 & 5037 & 2.53 & 5050 & 2.60 & 1.51 & 2.20 & 2.16  \\
 NGC 2168 &   81 & 4551 & 4483 & 4467 & 1.88 & 4500 & 1.65 & 1.78 & 2.81 & 3.01  \\
          &  310 & 5106 & 5316 & 5078 & 1.74 & 5150 & 1.10 & 2.76 & 3.16 & 3.78  \\
          &  662 & 4749 & 4521 & 4400 & 1.99 & 4500 & 1.80 & 1.80 & 2.78 & 2.84  \\
 NGC 6475 &HD~162587 & 4718 & 4631 & 4341 & 1.99 & $\ldots$ & $\ldots$ & $\ldots$ & $\ldots$ & $\ldots$ \\ 
  & HD~162587$^{1}$ & $\ldots$ & $\ldots$ & $\ldots$ & $\ldots$ & 4800 & 2.50 & 0.12 & $\ldots$ & $\ldots$ \\ 
  & HD~162587$^{2}$ & $\ldots$ & $\ldots$ & $\ldots$ & $\ldots$ & 5100 & 2.80 & 0.15 & $\ldots$ & $\ldots$ \\ 
     & HD~162391 & 4793 & 4875 & 4243 & 2.14 & 4900 & 2.00 & 2.07 & 2.56 & 2.49  \\
          &HD~162496 & 4595 & 4546 & 4112 & 2.10 & 4600 & 1.90 & 1.81 & 2.52 & 2.53  \\
 NGC 6991 &   22 & 5205 & 4842 & 4837 & 3.40 & 5300 & 3.45 & 1.14 & 1.18 & 1.00  \\
          &   67 & 4940 & 4843 & 4751 & 2.56 & 4950 & 2.80 & 1.45 & 1.93 & 1.62  \\
          &  100 & 4899 & 5092 & 4865 & 2.74 & 5050 & 2.90 & 1.27 & 1.74 & 1.56  \\
          &  131 & 4920 & 5161 & 5009 & 2.65 & 5050 & 2.90 & 1.35 & 1.83 & 1.61  \\
 NGC 7762 &   35 & 4630 & 4332 & 4629 & 2.22 & 4425 & 2.00 & 1.33 & 2.02 & 2.23  \\
          &   91 & 4595 & 4589 & 4722 & 2.22 & 4775 & 2.30 & 1.45 & 2.00 & 1.97  \\
          &  110 & 4662 & 4652 & 4863 & 2.60 & 4800 & 2.50 & 1.34 & 1.65 & 1.82  \\

\hline
\end{tabular}
\end{minipage}
\end{table*}

\section{Stellar parameters and chemical composition}
\subsection{Line list}

The line list comprising the atomic line data was taken from our previous papers (Reddy et al. 2012, 2013, 2015) and the line equivalent widths were measured  interactively from the spectra of program stars using the {\it splot} task in {\scs IRAF}. 
Our spectral line list of 23 elements (Na, Mg, Al, Si, Ca, Sc, Ti, V, Cr, Mn, Fe, Co, Ni, Cu, Zn, Y, Zr, Ba, La, Ce, Nd, Sm, and Eu) includes clean, unblended, isolated and symmetric spectral features within the spectral range 4450$-$8850 \AA. The high S/N ratios of the spectra helped in identifying the continuum regions and the spectrum normalization has been done more securely around the selected lines, their EWs can be measured with fair certainty.
 
Lines from  wavelength intervals with uncertain continuum placement due to heavy line crowding and/or affected by telluric contamination were excluded. Weak (EW$<$8 m\AA) as well as strong (EW$>$140 m\AA) lines were also discarded from the analysis. Chemical abundances for most elements well represented by lines are based on lines weaker than 120 m\AA, but strong lines were employed for species represented by a few lines (for example, Ba {\scs II}). Excluding the sub-giant star NGC 6991\#22, the EWs of the barium lines employed for abundance analysis cover the range 110$-$300 m\AA\ with a mean value of about 179$\pm$59 m\AA (16 giants). Our final list of 350 absorption lines per star contains on average  150 Fe {\scs I} lines occupying a range of $\sim$\,0.1 to 5.0 eV in lower excitation potential (LEP) and 20$-$140 m\AA\ in EWs, and 16 Fe {\scs II} lines with LEPs of about 2.8 to 3.9 eV and EWs from $\simeq$ 25 to 110 m\AA. 

\subsection{Stellar parameters} \label{sp_stars}

To determine the stellar parameters and the chemical abundances of program stars, we followed the standard spectroscopic technique that requires a line list, model photospheres and a spectral analysis code. We used the grid of ATLAS9 one-dimensional, line-blanketed plane-parallel uniform LTE models computed with updated opacity distribution functions from Castelli \& Kurucz (2003). The desired model photosphere characterized by a specific combination of temperature, gravity, microturbulence and metallicity was extracted from the extensive grid of ATLAS9 models via the linear interpolation software written by Carlos Allende Prieto\footnote{\url{http://www.as.utexas.edu/~hebe/stools/}}. To compute the chemical abundances, we used the line list and model photosphere as inputs to the LTE line analysis and spectrum synthesis code {\scs \bf MOOG}\footnote{{\scs \bf MOOG} was developed and updated by Chris Sneden and originally described in Sneden (1973)}. 

However, the first step in the abundance analysis is the selection of a suitable model photospheric temperature and gravity. We obtained such preliminary estimates using the optical and 2MASS photometric colours (B-V), (V-K$_{\rm s}$) and (J-K$_{\rm s}$) following the precepts discussed in Reddy et al. (2012). We derived the star's photometric effective temperature, T$^{\rm phot}_{\rm eff_\star}$, by substituting the dereddened\footnote{The adopted interstellar extinctions are (A$_{V}$, A$_{K}$, E(V-K), E(J-K))= (3.1, 0.28, 2.75, 0.54)*E(B-V), where E(B-V) is taken from the {\scs WEBDA} database} photometric colours into the infrared flux method based colour$-$temperature calibrations of Alonso, Arribas \& Mart\'{i}nez-Roger (1999).

The photometric estimates of surface gravities, $\log~g_{\rm phot}$, were made by incorporating the heliocentric distance of the cluster, photometric temperature, bolometric correction $BC_{V}$, cluster turn-off mass $M_{\star}$ and the solar parameters of T$_{\rm eff},_{\odot}$= 5777 K and log~$g_{\odot}$= 4.44 cm s$^{-2}$ into the well known log~$g$\,$-$\,T$_{\rm eff}$ relation (Reddy et al. 2012). Estimates of $BC_{V}$s were made using Alonso et al.'s (1999) calibrations connecting the photometric temperature and metallicity. 
 
The turn-off masses of giants have been estimated by achieving a good match between the cluster CMD and Padova stellar evolutionary tracks of Marigo et al. (2008): the adopted turn-off masses are 3.6, 4.6, 3.8, 2.1 and 1.5 $M_{\odot}$ for Stock 2, NGC 2168, NGC 6475, NGC 6991 and NGC 7762, respectively. Although the initial set of fundamental parameters (age, distance, reddening and metallicity) of OCs utilised in colour$-$temperature calibrations and isochrone fitting is drawn from the {\scs WEBDA} database drawn, they were revised later using the spectroscopic estimates of cluster metallicities.

\begin{figure*}
\begin{center}
\includegraphics[trim=0.1cm 3.1cm 5.5cm 4.4cm, clip=true,width=0.85\textwidth,height=0.47\textheight]{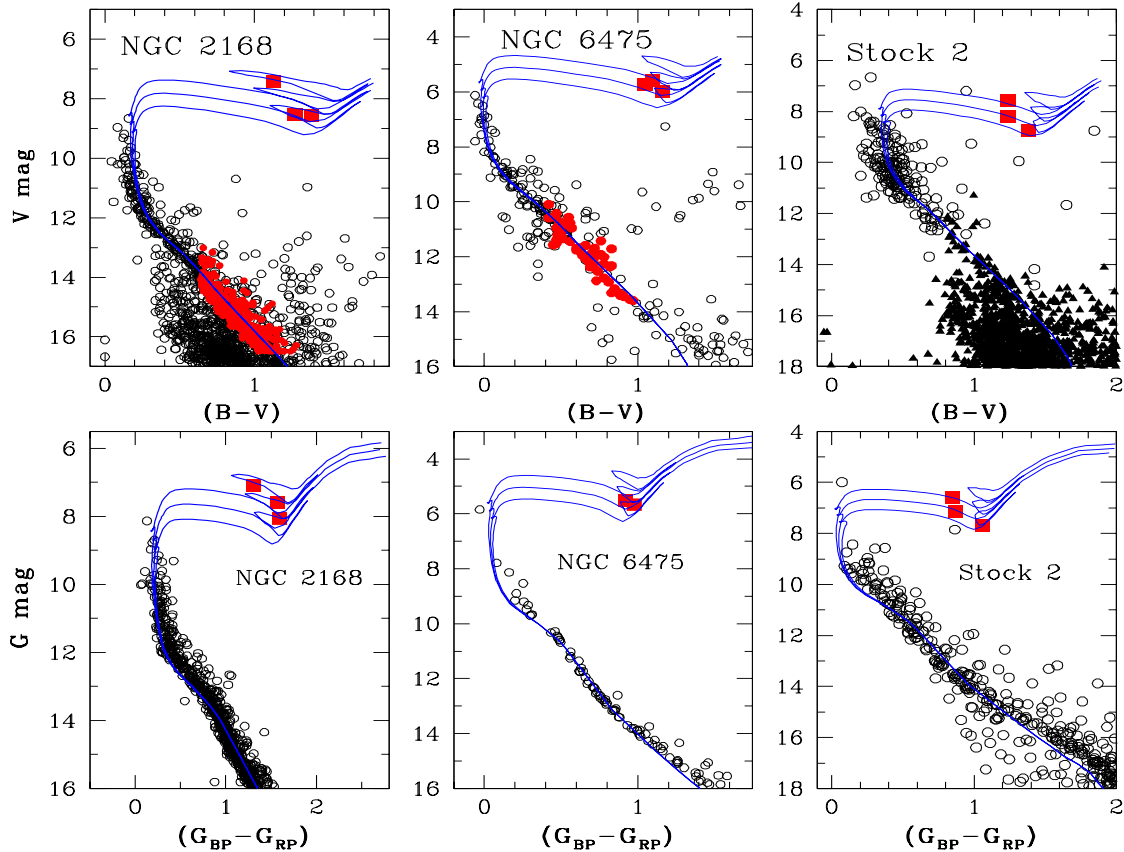} \vspace{-0.15cm}
\caption{CMDs of OCs NGC 2168, NGC 6475, and Stock 2 based on BV (top row) and Gaia DR2 photometry (bottom row). Top row: The radial velocity and proper motions confirmed members of each cluster are highlighted with red filled circles (dwarfs) and squares (giants). Bottom row: Common proper motions and distance members using Gaia DR2 photometry with giants analysed in this paper represented by the red filled squares. In top and bottom rows, the isochrones in blue are constructed for [Fe/H]=$-$0.1 dex (NGC 2168), $-$0.01 dex (NGC 6475), and $-$0.06 dex (Stock 2) and are shown for ages of log$(Age)=$\,8.1$\pm$0.1 yr (NGC 2168), 8.2$\pm$0.1 yr (NGC 6475), and 8.35$\pm$0.1 yr (Stock 2) (see table \ref{rev_param}). }
\label{cmdcluster}
\end{center}
\end{figure*}

To mitigate the systematic errors in our final stellar abundances, we performed a differential abundance analysis of stars relative to the Sun by running the {\it abfind} driver of {\scs \bf MOOG}. Following Reddy et al. (2012), reference solar abundances were derived using solar EWs, measured off the solar integrated disk spectrum (Kurucz et al. 1984), and adopting the ODFNEW grid of Kurucz model photosphere with T$_{\rm eff},_{\odot}$ = 5777 K, log~$g_{\odot}$ = 4.44 cm s$^{-2}$ and [Fe/H]$=$0.00. We found a microturbulence velocity of $\xi_{t}$ = 0.93 km s$^{-1}$ using Fe {\scs II} lines. 

Spectroscopic atmospheric parameters (i.e., effective temperature T$_{\rm eff}$, surface gravity log~$g$, microturbulence $\xi_{t}$ and metallicity [Fe/H]) of program stars were derived in an iterative manner using iron line EWs thanks to numerous Fe {\scs I} lines with good coverage in line's LEP from $\sim$ 0.0 to 5.0 eV throughout the optical region of spectra as well as Fe {\scs II} lines with a range in measured EWs. Starting with a Kurucz model with photometric estimates of temperature and gravity, the individual iron line abundances were force-fitted to match the model-generated iron line EWs to observed ones by imposing the conditions of excitation and ionization balance of Fe {\scs I} and Fe {\scs II} lines as well as the independence between the iron abundances and line's reduced EWs.
First, the microturbulence assumed to be isotropic and depth independent was derived by requiring that the iron abundance from Fe {\scs I} and Fe {\scs II} lines be independent of a line's reduced equivalent width, REW$=$log~(EW/$\lambda$). The value of $\xi_{t}$ derived from iron lines is confirmed by other species due to Ni {\scs I}, Ti {\scs I}, Ti {\scs II}, V {\scs I}, Cr {\scs I} and Cr {\scs II}.   
Second, the effective temperature was adjusted in steps of 25 K until the slope of iron abundance from Fe {\scs I} lines with line's LEP was less than 0.004 dex/$eV$ (excitation equilibrium). This condition is also reasonably satisfied for the lines of other species like Ti {\scs I} and Ni {\scs I}. Third, the surface gravity is adjusted until the difference in average abundances of Fe {\scs I} and Fe {\scs II} lines is smaller than 0.02 dex for the derived T$_{\rm eff}$ and $\xi_{t}$ (i.e. ionization balance between the neutral and ionized species). Ionization balance is also satisfied reasonably well by other species such as Ti and Cr having neutral and ionized lines.
The final stellar parameters provided in Table \ref{stellar_param} were obtained via several iterations when these three conditions were simultaneously satisfied.

Following the procedure described in Reddy \& Lambert (2015), we obtained uncertainties of 50 K, 0.1 dex and 0.1 km s$^{-1}$ in $T_{\rm eff}$, log~$g$ and $\xi_{t}$, respectively, where each stellar parameter is varied while keeping other two parameters fixed until the three conditions imposed in derivation of stellar parameters produce spurious slopes and introduce $\pm$1$\sigma$ shift each in the average Fe abundance and in the difference between Fe {\scs I} and Fe {\scs II} abundances.

A comparison of photometric atmospheric parameters with estimates from spectroscopy is offered in table \ref{stellar_param}. With few exceptions, our spectroscopic $T_{\rm eff}$s and log~$g$s are in good agreement with photometric ones: Mean differences in photometric temperatures estimated using (B-V) and (V-K) is $+$9 $\pm$ 184 K and using (V-K) and (J-K) is $+$95 $\pm$ 250 K. The corresponding mean differences between (B-V), (V-K) and (J-K) based colour temperatures and the spectroscopic T$^{spec}_{\rm eff}$'s are $-$9 $\pm$ 128 K, $-$19 $\pm$ 201 K and $-$114 $\pm$ 218 K, respectively. No single color-based temperature agrees well with corresponding spectroscopic temperatures for all stars. However, the differences between photometric and spectrographic estimates are within the uncertainties found between different colour-based temperatures for the same star. The mean differences in gravities and luminosities across the sample of 17 stars (neglecting HD~162587) are $+$0.14$\pm$0.26 dex and $-$0.10$\pm$0.27, respectively.

The photometric stellar parameters derived from the infrared flux method based colour$-$temperature calibrations are mainly sensitive to the adopted colours, reddening, metallicity and distance modulus of the cluster. An uncertainty of 0.02 mag. each in (B-V) and E(B-V), and an error of 0.05 dex in metallicity translates to errors of 65 K and 52 K in temperatures measured from (B-V) and (V-K) relations, respectively. Although the (J-K) vs. T$_{\rm eff}$ relation is independent of metallicity, errors of 0.02 mag. and 0.011 (i.e., 0.54*E(B-V)), respectively in colour and reddening contributes a total temperature uncertainty of 68 K. Here the total error in T$_{\rm eff}$ is the quadratic sum of errors introduced by varying the respective parameters used in the photometric relations. Note, however, that these are the lower limits whose values tend to increase if true errors are adopted. Similarly, the above errors in T$_{\rm eff}$s and an uncertainty of 0.2 mag. in distance modulus yield an uncertainty of 0.08 dex in photometric estimates of surface gravities. Again the inclusion of differential reddening instead of adopting a single value for all members in a given cluster may further boost uncertainties in photometric stellar parameters.

These errors in photometric stellar parameters are too large given the small uncertainties in the reddening estimates. Should we adopt the photometric estimates for model photospheres in the abundance analysis, we will find a steep slope of Fe {\scs I} abundance with line's LEP and the ionization balance between the neutral and ionized species will also be disturbed. Therefore, to be consistent, we will consider the spectroscopic model photospheres as the final atmospheric parameters for stars under analysis.

The measurement of chemical abundances of program stars was extended to other species using line EWs and the spectroscopic atmospheric parameters derived previously from iron line EWs. But synthetic profiles were computed and matched to the stellar spectra to derive abundances for lines affected by hyperfine structure (hfs) and isotopic shifts and/or affected by blends. The suite of lines included in the synthetic spectrum analysis are Sc, Mn, Cu, Zn, Ba and Eu. We followed the standard procedure of synthetic profile fitting by running the {\it synth} driver of {\scs \bf MOOG}.

The chemical abundances for the individual cluster members averaged over all available lines of given species are presented in Tables \ref{abu_stock2}$-$\ref{abu_ngc7762}, relative to solar abundances derived from the adopted $gf$-values (see Table 4 from Reddy et al. 2012). Table entries provide the average [Fe/H] and [El/Fe] for all elements along with the number of lines used in calculating the abundance of that element. Within the run from Na to Eu, all stars in a given cluster have very similar [El/Fe] for almost all the elements. 

To compute the errors in elemental abundances, we repeated the abundance analysis by varying each stellar parameter separately by an amount equal to its uncertainty, while keeping other two parameters fixed. Additionally, we considered the error introduced by varying the model metallicity by 0.1 dex. The systematic uncertainty associated with the abundances is the quadratic sum of these four error terms. The random error in the mean abundance of a species El is replaced by the $\sigma_{[El/Fe]}$/$\sqrt(N_{\rm lines})$, where N$_{\rm lines}$ is the number of lines of species El. The total error $\sigma_{tot}$ for each of the species is the quadratic sum of the systematic and random errors. The final mean chemical composition of each OC and the $\sigma_{tot}$ from this study are presented in Table \ref{mean_abundance}.

Inspection of Tables \ref{abu_stock2}$-$\ref{abu_ngc7762} shows that the giants common to a cluster have the same composition to within the estimated total error with almost no exceptions. The exceptions are abundances (e.g., Cu) based on a single line. Although the number of giants per cluster is few, their common composition within a cluster is consistent with the assumption of a chemical homogeneity for a cluster. Each giant was analyzed independently of all other giants and without knowledge about its host cluster and, thus, abundances for giants belonging to a common cluster provide an estimate of the measurement uncertainties. 

Examination of Table \ref{mean_abundance} shows, as anticipated, for clusters with [Fe/H] $\simeq 0.0$, that [El/Fe] $\simeq 0.0$ for almost every element lighter than Ni in every cluster. Sodium might appear to be an exception but Na, as discussed below, is enriched through the first dredge-up. Ba to Eu enrichments in some clusters are expected, as discussed below (Lambert \& Reddy 2016). Possible exceptions  include Cu and Zn with [Cu/Fe] and [Zn/Fe] at $-0.2$ in NGC 6475 but Cu and Zn abundances are based on a single line each. In the Mg - Ni group, the only obvious outstanding [El/Fe] entries in Table \ref{mean_abundance} are those for Si for NGC 2168 and NGC 7762.  Their entries of [Si/Fe] of $+0.28$ and  $+0.23$, respectively, are shared by each giant in the cluster. These entries seem odd from the nucleosynthetic point of view because  the other $\alpha$-elements  (Mg, Ca and Ti) do not share this apparent overabundance.

\begin{figure}
\begin{center}
\includegraphics[trim=0.4cm 3.8cm 9.4cm 4.3cm, clip=true,width=0.52\textwidth,height=0.4\textheight]{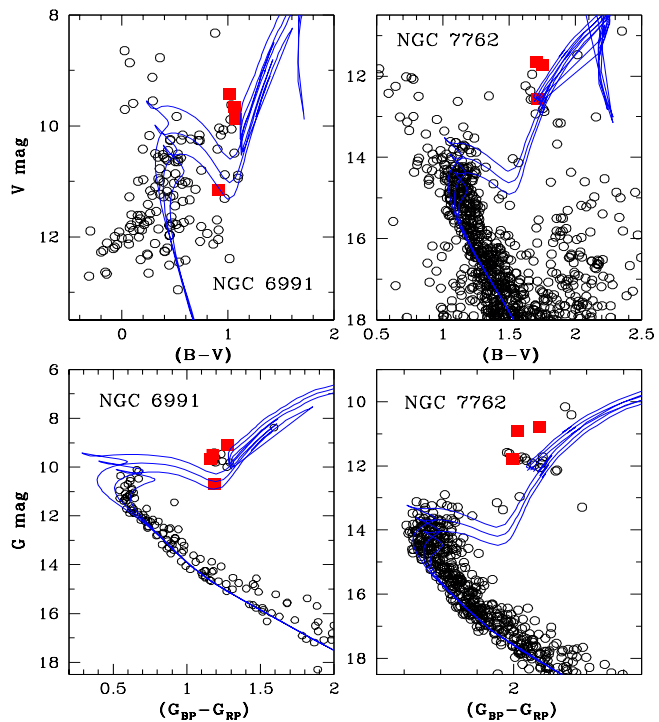} \vspace{-0.15cm}
\caption[]{Same as Figure \ref{cmdcluster}, but for the OCs NGC 6991 and NGC 7762. Isochrones in blue are constructed for [Fe/H]=\,0.0 dex and log$(Age)=$\,9.0$\pm$0.1 yr (NGC 6991) and [Fe/H]=$-$0.07 dex and log$(Age)=$\,9.4$\pm$0.1 yr (NGC 7762). Stars observed for abundance analysis in this study are marked with red filled squares. }  
\label{cmd6991}
\end{center}
\end{figure}

\begin{table*}
\caption{Elemental abundance ratios [El/Fe] for elements from Na to Eu for Stock 2, NGC 2168, 6475, 6991 and 7762 from this study. Abundances calculated by synthesis are presented in bold typeface.}
\label{mean_abundance}
\begin{tabular}{lcccccc}   \hline
 \multicolumn{1}{c}{Species} & \multicolumn{1}{c}{Stock 2} & \multicolumn{1}{c}{NGC 2168} & \multicolumn{1}{c}{NGC 6475} &
 \multicolumn{1}{c}{NGC 6991} & \multicolumn{1}{c}{NGC 7762}   \\ \hline

$[$Na I/Fe$]$	  & $+0.27\pm0.03$   & $+0.22\pm0.03$   & $+0.23\pm0.03$    & $+0.10\pm0.03$   & $+0.11\pm0.03$  \\
$[$Mg I/Fe$]$	  & $+0.03\pm0.03$   & $+0.06\pm0.04$   & $+0.01\pm0.03$    & $-0.07\pm0.03$   & $+0.03\pm0.03$  \\
$[$Al I/Fe$]$	  & $+0.06\pm0.04$   & $+0.03\pm0.03$   & $+0.04\pm0.03$    & $+0.08\pm0.03$   & $+0.05\pm0.03$  \\
$[$Si I/Fe$]$	  & $+0.09\pm0.04$   & $+0.28\pm0.05$   & $+0.06\pm0.05$    & $+0.04\pm0.05$   & $+0.23\pm0.05$  \\
$[$Ca I/Fe$]$	  & $-0.04\pm0.03$   & $+0.01\pm0.04$   & $-0.03\pm0.03$    & $+0.01\pm0.03$   & $-0.02\pm0.03$  \\
$[$Sc I/Fe$]$	  & $+0.03\pm0.02$   &    $\ldots$      & $\,0.00\pm0.05$   & $+0.01\pm0.04$   &    $\ldots$      \\
$[$Sc II/Fe$]$	  & $+0.02\pm0.04$   & $-0.04\pm0.03$   & $-0.01\pm0.03$    & $-0.02\pm0.03$   & $\,0.00\pm0.04$  \\
$[$Sc II/Fe$]$	  &$\bf+0.06\pm0.02$ &$\bf-0.04\pm0.02$ &$\bf-0.04\pm0.03$  &$\bf+0.01\pm0.02$ &$\bf+0.07\pm0.02$  \\
$[$Ti I/Fe$]$	  & $-0.06\pm0.04$   & $-0.12\pm0.04$   & $-0.02\pm0.04$    & $-0.01\pm0.03$   & $-0.06\pm0.04$  \\
$[$Ti II/Fe$]$	  & $-0.08\pm0.03$   & $-0.11\pm0.03$   & $-0.04\pm0.03$    & $-0.06\pm0.03$   & $-0.06\pm0.04$  \\
$[$V I/Fe$]$	  & $-0.03\pm0.04$   & $-0.05\pm0.05$   & $-0.01\pm0.05$    & $-0.03\pm0.04$   & $+0.01\pm0.05$  \\
$[$Cr I/Fe$]$	  & $\,0.00\pm0.03$  & $-0.02\pm0.03$   & $+0.01\pm0.03$    & $-0.02\pm0.03$   & $+0.01\pm0.03$  \\
$[$Cr II/Fe$]$	  & $+0.03\pm0.03$   & $\,0.00\pm0.03$  & $+0.02\pm0.03$    & $+0.02\pm0.02$   & $+0.07\pm0.04$  \\
$[$Mn I/Fe$]$	  &$\bf-0.07\pm0.02$ &$\bf-0.06\pm0.03$ &$\bf-0.06\pm0.02$  &$\bf-0.01\pm0.02$ &$\bf-0.13\pm0.02$  \\
$[$Fe I/H$]$	  & $-0.06\pm0.04$   & $-0.11\pm0.04$   & $-0.01\pm0.04$    & $+0.01\pm0.04$   & $-0.07\pm0.04$  \\
$[$Fe II/H$]$	  & $-0.06\pm0.05$   & $-0.11\pm0.04$   & $-0.01\pm0.05$    & $\,0.00\pm0.05$  & $-0.08\pm0.05$  \\
$[$Co I/Fe$]$	  & $-0.01\pm0.02$   & $+0.03\pm0.02$   & $+0.01\pm0.03$    & $+0.01\pm0.03$   & $+0.11\pm0.03$  \\
$[$Ni I/Fe$]$	  & $-0.03\pm0.04$   & $-0.02\pm0.04$   & $-0.04\pm0.04$    & $-0.01\pm0.03$   & $+0.01\pm0.04$  \\
$[$Cu I/Fe$]$	  &$\bf-0.11\pm0.01$ &$\bf-0.11\pm0.02$ &$\bf-0.17\pm0.02$  &$\bf-0.03\pm0.01$ &$\bf+0.03\pm0.01$  \\
$[$Zn I/Fe$]$	  &$\bf-0.03\pm0.04$ &$\bf-0.14\pm0.04$ &$\bf-0.22\pm0.04$  &$\bf+0.01\pm0.04$ &$\bf-0.07\pm0.04$  \\
$[$Y II/Fe$]$	  & $-0.01\pm0.03$   & $-0.01\pm0.03$   & $\,0.00\pm0.03$   & $+0.09\pm0.04$   & $+0.04\pm0.04$  \\
$[$Zr I/Fe$]$	  & $+0.03\pm0.05$   & $-0.11\pm0.05$   & $+0.04\pm0.04$    & $+0.01\pm0.05$   & $+0.01\pm0.04$  \\
$[$Zr II/Fe$]$	  & $+0.03\pm0.04$   & $+0.18\pm0.05$   & $+0.07\pm0.04$    & $+0.09\pm0.04$   & $+0.03\pm0.04$  \\
$[$Ba II/Fe$]$	  &$\bf+0.26\pm0.05$ &$\bf+0.17\pm0.05$ &$\bf+0.12\pm0.05$  &$\bf+0.12\pm0.05$ &$\bf+0.07\pm0.05$  \\
$[$La II/Fe$]$	  & $-0.06\pm0.04$   & $-0.10\pm0.04$   & $-0.03\pm0.05$    & $+0.12\pm0.04$   & $+0.05\pm0.05$  \\
$[$Ce II/Fe$]$	  & $+0.04\pm0.04$   & $+0.07\pm0.03$   & $+0.01\pm0.04$    & $+0.17\pm0.04$   & $+0.05\pm0.04$  \\
$[$Nd II/Fe$]$	  & $+0.08\pm0.04$   & $+0.07\pm0.04$   & $+0.06\pm0.04$    & $+0.19\pm0.04$   & $+0.08\pm0.04$  \\
$[$Sm II/Fe$]$	  & $+0.09\pm0.04$   & $+0.03\pm0.04$   & $+0.03\pm0.04$    & $+0.24\pm0.05$   & $+0.09\pm0.05$  \\
$[$Eu II/Fe$]$	 &$\bf+0.09\pm0.03$  &$\bf+0.03\pm0.03$ &$\bf+0.10\pm0.03$  &$\bf+0.15\pm0.03$ &$\bf+0.08\pm0.03$  \\

\hline
\end{tabular}
\end{table*}

\section{Revised cluster parameters}

With the membership information based on proper motions and radial velocities of stars and the spectroscopic values of [Fe/H] measured to an accuracy of 0.05 dex, we have determined a refined set of cluster fundamental parameters using the Padova suite of stellar evolutionary tracks (Marigo et al. 2008). 

To construct the CMDs, we relied mostly on the BV and Gaia DR2 photometry for all but the OC NGC 6991. For the cluster NGC 6991 whose (V, B-V) CMD is poorly populated and clearly lacks a well-defined cluster main-sequence, we measured the cluster parameters from Gaia DR2 CMD. Sources of the cluster UBV data include Krzeminski \& Serkowski (1967) and Foster et al. (2000) for Stock 2, Sung \& Bessell (1999; NGC 2168), Prosser et al. (1995; NGC 6475), Kharchenko et al. (2005; NGC 6991) and Maciejewski et al. (2008; NGC 7762). Figure \ref{cmdcluster} provides (V, B-V) CMDs for all but clusters NGC 6991  and NGC 7762 with the radial velocity and proper motions confirmed members of each cluster highlighted with red filled circles (dwarfs) and squares (giants). The CMD of NGC 6991 and NGC 7762 will be discussed separately in the next section. Sources of proper motions and RVs of stars in each of the OCs are discussed previously in Section \ref{rv_mem}. 

\begin{figure*}
\begin{center}
\includegraphics[trim=0.1cm 7.0cm 3.5cm 4.35cm, clip=true,width=0.85\textwidth,height=0.25\textheight]{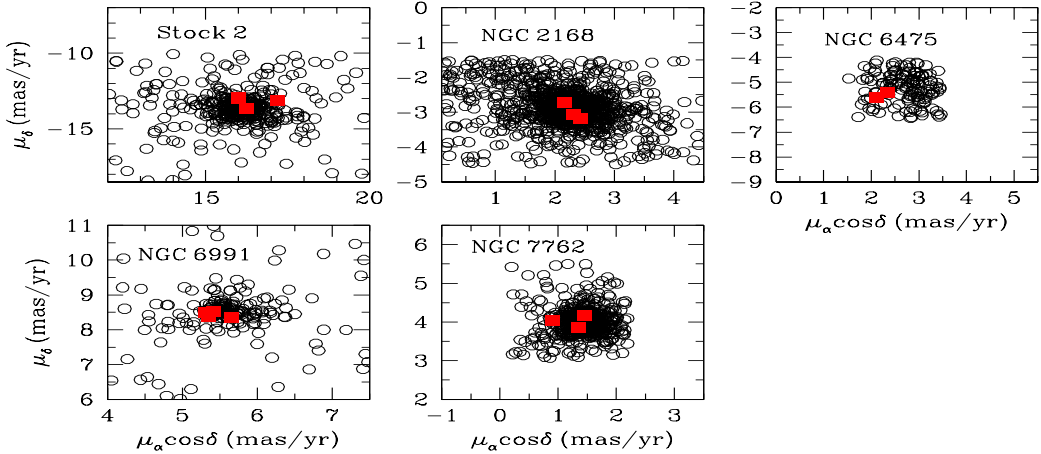}
\caption[]{ The individual proper motions of the stars used in constructing the CMDs based on Gaia photometry (Figures \ref{cmdcluster} \& \ref{cmd6991}). Stars marked as red filled squares are the red giants analysed in this paper. }  
\label{pmcluster}
\end{center}
\end{figure*}

We generated several Padova isochrones of varying age with a fixed spectroscopic [Fe/H] and compared them with the observed CMDs (Figure \ref{cmdcluster}). Starting with the published cluster age, distance and reddening, the isochrones were shifted horizontally along the colour axis to determine the reddening E(B-V) and vertically to deduce the distance modulus (m-M)$_{\rm V}$ until the MS turn-off point, MS slope and colour, and the red clump were fitted simultaneously. Finally, we retained in each CMD a set of three isochrones separated in age by 0.1 dex that best covers much of the lower main-sequence down to fainter magnitudes. The best-fitting isochrone was judged by visual inspection of CMDs in figure \ref{cmdcluster}, where the middle isochrone correspond to cluster's age with an error of 0.1 dex represented by the neighbouring tracks. We estimated the reddening and distance modulus of the cluster from the best fitting evolutionary track with typical errors of 0.04 and 0.15 dex, respectively.

In general the isochrone fits are reasonably good for the main-sequence, turn-off, and red clump regions of all OCs. Although the red clump colour in the observed CMD of Stock 2 is bluer by 0.04 than in the model evolutionary tracks, such a small difference is comparable to systematic errors in the calibration of the photometry. We have verified the cluster parameters obtained using the BV photometry with those calculated from  Gaia astrometry (Gaia Collaboration et al. 2018). The CMDs of OCs based on Gaia DR2 photometry with the set of theoretical isochrones from Marigo et al. (2008) are shown in the bottom rows of Figure \ref{cmdcluster} and Figure \ref{cmd6991}. The proper motions of cluster members used in constructing these Gaia CMDs are shown in figure \ref{pmcluster}. Theoretical isochrones from the isochrone tables are transformed from the Johnson-Cousins photometric system to Gaia photometry using the transformation polynomials involving V, (V-I), G and (G$_{BP}$-G$_{RP}$) (Jordi et al. 2010). These transformed isochrones best fit the CMDs based on Gaia photometry for the cluster parameters listed in table \ref{rev_param}. Further refinement of cluster parameters using isochrones is not necessary for this paper which focuses mainly on the detailed chemical composition of each cluster.

We present in Table \ref{rev_param} the revised fundamental parameters of OCs from this study including those values calculated using the Gaia astrometry. An excellent agreement between the cluster parameters obtained independently using the BV photometry and Gaia DR2 photometry and astrometry is evident. A close match of the astrometric parameters of red giants analysed in this paper (Table \ref{cluster_redgiants}) with cluster parameters obtained from CMDs and the cluster mean parameters from Gaia DR2 (Table \ref{rev_param}) confirm the membership of red giants to OCs. Thus, the Gaia DR2 astrometry and photometry provides an independent measure of the cluster fundamental parameters, including the improved proper motions, distances and membership information of cluster giants analysed in this work.

\begin{table*}
\centering
\caption{Cluster fundamental parameters derived in this study using the BV (Columns 3$-$6) and Gaia based CMDs (Columns 7$-$10). The cluster parameters derived using the Gaia CMDs are in excellent agreement with those determined from CMDs based on BV photometry. }
\label{rev_param}
\begin{tabular}{lccccccccc}   \hline
\multicolumn{1}{l}{Cluster} & \multicolumn{1}{c}{[Fe/H]} & \multicolumn{1}{c}{E(B-V)} & \multicolumn{1}{c}{V$-$M$_{\rm V}$} & \multicolumn{1}{c}{Age} & \multicolumn{1}{c}{d$_{BV}$ } & \multicolumn{1}{c}{E(G$_{BP}$-G$_{RP}$)} & DM$_{G}$ & \multicolumn{1}{c}{d$_{G}$} & \multicolumn{1}{c}{Age}  \\ 
\multicolumn{1}{l}{ } & \multicolumn{1}{c}{(dex)} & \multicolumn{1}{c}{ } & \multicolumn{1}{c}{ } & \multicolumn{1}{c}{(Myr)} & \multicolumn{1}{c}{(pc)} & \multicolumn{1}{c}{ } & \multicolumn{1}{c}{ } & \multicolumn{1}{c}{(pc) } & \multicolumn{1}{c}{(Myr)}  \\    
\hline
 Stock 2   & $-$0.06  &  0.45  & 09.35 &  225 & 390 & 0.58 & 09.05 & 372 &  225  \\
 NGC 2168  & $-$0.11  &  0.31  & 10.65 &  125 & 866 & 0.40 & 10.50 & 861 &  125  \\
 NGC 6475  & $-$0.01  &  0.10  & 07.50 &  158 & 274 & 0.13 & 07.45 & 273 &  158  \\
 NGC 6991  & $\,$0.00 &  0.12  & 09.20 & 1000 & 582 & 0.15 & 09.10 & 573 & 1000  \\
 NGC 7762  & $-$0.07  &  0.63  & 11.75 & 2500 & 910 & 0.81 & 11.45 & 904 & 2500  \\ 
 \hline
 \end{tabular} 
\end{table*}

\begin{table*}
\centering
\caption{Cluster fundamental parameters derived by taking the mean of Gaia astrometric parameters of cluster members (Columns 2$-$5). Columns 6$-$9 lists the mean distance, parallax and proper motions of red giants analysed in this paper. An excellent agreement of astrometric parameters of red giants with cluster parameters confirm the membership of red giants to OCs. } 
\label{cluster_redgiants}
\begin{tabular}{lcccc|cccccc}   \hline
\multicolumn{5}{c}{Cluster mean parameters derived using all cluster stars } \vline & \multicolumn{4}{c}{Mean of red giants analysed in this paper}  \\
\multicolumn{1}{c}{Cluster} & \multicolumn{1}{c}{d$_{\rm Gaia}$} & \multicolumn{1}{c}{Plx$_{\rm Gaia}$} & \multicolumn{1}{c}{pmRA$_{\rm Gaia}$} &  \multicolumn{1}{c}{pmDEC$_{\rm Gaia}$} \vline & \multicolumn{1}{c}{d$_{\rm Gaia}$} & \multicolumn{1}{c}{Plx$_{\rm Gaia}$} & \multicolumn{1}{c}{pmRA$_{\rm Gaia}$} & \multicolumn{1}{c}{pmDEC$_{\rm Gaia}$}    \\
\multicolumn{1}{c}{} & \multicolumn{1}{c}{(pc)} & \multicolumn{1}{c}{(mas)} & \multicolumn{1}{c}{(mas/yr)} & \multicolumn{1}{c}{(mas/yr)} \vline & \multicolumn{1}{c}{(pc)} & \multicolumn{1}{c}{(mas)} & \multicolumn{1}{c}{(mas/yr)} & \multicolumn{1}{c}{(mas/yr)}   \\ 
\hline
Stock 2   & 387$^{+130}_{-78}$& $2.58\pm0.65$ & $16.148\pm0.646$ & $-13.680\pm0.600$ & 375$^{+4}_{-4}$  & $2.66\pm0.03$ & $16.547\pm0.619$ & $-13.250\pm0.310$      \\
NGC 2168  & 869$^{+65}_{-56}$ & $1.15\pm0.08$ & $2.279\pm0.227$  & $-2.920\pm0.297$ & 877$^{+48}_{-43}$ & $1.14\pm0.06$ & $2.300\pm0.107$  & $-2.993\pm0.189$      \\
NGC 6475  & 274$^{+11}_{-10}$ & $3.65\pm0.14$ & $2.535\pm0.346$  & $-5.004\pm0.464$ & 273$^{+6}_{-6}$  & $3.66\pm0.08$ & $2.224\pm0.122$  & $-5.519\pm0.108$      \\
NGC 6991  & 562$^{+55}_{-46}$ & $1.78\pm0.16$ & $5.584\pm0.707$  & $8.491\pm0.719$ & 561$^{+6}_{-6}$   & $1.78\pm0.02$ & $5.435\pm0.134$  & $8.433\pm0.073$       \\
NGC 7762  & 990$^{+62}_{-55}$ & $1.01\pm0.06$ & $1.439\pm0.277$  & $4.033\pm0.448$ & 990$^{+10}_{-10}$ & $1.01\pm0.01$ & $1.237\pm0.235$  & $4.020\pm0.126$      \\

\hline
\end{tabular} 
\flushleft
\end{table*}

\subsection{NGC 6991}
Kharchenko et al. (2005) performed the first and only photometric membership study of the poorly populated OC NGC 6991 whose BV photometry collected from the All-Sky Compiled Catalogue of 2.5 Million Stars (ASCC-2.5, Kharchenko et al. 2001) covers a 24 arcminute radius field of view centered on the cluster at $\alpha({\rm J2000})$=20$^{\rm h}$ 54$^{\rm m}$ 32$^{\rm s}$, $\delta({\rm J2000})=+$47${\degr}$27${\arcmin}$. Yet the majority of faint stars in the cluster's field of view lack UBV data due to the relatively bright limiting magnitude (V$\approx$ 14 mag) of the ASCC-2.5 catalogue. Kharchenko et al. estimated a cluster age of 1.3 Gyr, a reddening of E(B-V)$=$0.0 and a heliocentric distance of 700 parsec by fitting isochrones to the sparsely populated (V, B-V) colour-magnitude diagram such as the one shown in Figure \ref{cmd6991}.
 
However, the ready availability of Gaia DR2 photometry and astrometry from Gaia mission (Gaia Collaboration et al. 2018) helped us to to derive accurate cluster properties. Gaia DR2 photometry was collected only for stars of common parallax and proper motions in a radius of 25-arcminute field around the cluster center. 

The first panel in the bottom row of Figure \ref{cmd6991} displays the Gaia based CMD constructed from the sample of stars whose individual proper motions have resulted a mean cluster motion of $\mu_{\alpha}$ cos $\delta=$+5.58$\pm$0.71 mas yr$^{-1}$ in right ascension and $\mu_{\delta}=$+8.49$\pm$0.72 mas yr$^{-1}$ in declination. Gaia CMD reveals a well-defined main-sequence along with a clump of red giants at G(mag)$\sim$10 and (G$_{BP}$-G$_{RP}$)$\sim$1.2.

The optimum fit by eye to the Gaia DR2 observations yields a reddening of E(G$_{BP}$-G$_{RP}$)$=$0.15$\pm$0.04 (i.e., E(B-V)$=$0.12$\pm$0.03) and a distance modulus of G$-$M$_{\rm G}=$\,9.1$\pm$0.1 (V$-$M$_{\rm V}=$\,9.2$\pm$0.1), corresponding to a cluster distance of d\,$=$\,573$\pm$25 parsec from the Sun. The solar metallicity Padova isochrone from Marigo et al. (2008) that best fits the observations has log\,(t)=9.0$\pm$0.1 yr that translates to a cluster age of 1.0$\pm$0.2 Gyr. We have verified that the cluster distance and age obtained from Gaia CMD also fits well the (J, J-H) and (K$_{\rm s}$, J-K$_{\rm s}$) CMDs for the JHK$_{s}$ photometry taken from 2MASS survey (Skrutskie et al. 2006).

\section{Results}

\subsection{Comparisons with the literature} 

An aim of our abundance analyses is to provide information on chemical compositions for open clusters not previously studied
spectroscopically. This aim often means that there is a scarcity of abundance determinations in the literature with which to compare
our results. At present, there are analyses from  spectra for giants in two (NGC 6475 and NGC 7762) of the five clusters in
the present paper and  iron abundance determinations for NGC 6691 and NGC 7762.

\subsection{NGC 6475}

The chemical content of NGC 6475 was determined previously by Villanova, Carraro \& Saviane (2009) from UVES high-resolution spectra (R = 80000) of the red giants HD 162391 and HD 162587 and two B-type  and three F-K-type main sequence stars. Here, we restrict the comparison to the red giant HD 162391; HD 162587 is rejected because we find it to be a double-lined spectroscopic binary. Systematic differences between the hot and cool main sequence stars and the giant HD 162391 may vitiate an abundance comparison.

For HD 162391, our [Fe/H] $= -0.01\pm0.03$ is in fine agreement with $+0.02\pm0.01$ determined by Villanova et al. Their abundance ratios [El/Fe]
agree with ours to $\pm0.10$ except for Na, V, Zn and Ba with the largest (Us - Villanova) of -0.29 occurring for Na.  Our analysis of HD 162391
is fully confirmed by our results for HD 162496 (Table \ref{abu_NGC2168_6475}).

Blanco-Cuaresma et al. (2015) provide an abundance analysis of a main sequence star in NGC 6475. Elements from Na to Ba were considered.
The [Fe/H] of $+0.04\pm0.05$ is in good agreement with our [Fe/H] $-0.01\pm0.03$ from the two red giants. Inspection of the differences (Us - B-C)  shows that for all the elements in common but Na they range from only $-0.07$ to $+0.06$. The Na difference is $+0.34$ indicating the red giant's expected Na enrichment from the first dredge-up (see below). A sceptic might question the validity of a comparison involving a dwarf and a giant on the grounds that systematic effects may
vitiate the comparison.  Such a question may be examined because Blanco-Cuaresma et al. analysed six clusters for which spectra of both dwarfs and giants were available. In all six clusters, the differences in [Fe/H] and [El/Fe] between giants and dwarfs (except, of course for Na) were within $\pm0.10$ and mean difference were within $\pm0.05$ dex except for Mg ($+0.06$), Si ($+0.10$) and Sc ($+0.07$) with standard
deviations of less than 0.03 except for Ba ($0.06$).  In brief, the above comparison for NGC 6475 of a dwarf with a giant appears valid.

\subsection{NGC 6991}

The [Fe/H] of NGC 6991 was determined from high-resolution spectra by Casamiquela et al. (2017): they in their Table 9 found [Fe/H] $= -0.01\pm0.03$
from six giants. Our value (Table \ref{mean_abundance}) from four giants is $+0.01\pm0.04$) in excellent agreement.  Casamiquela et al. report [Fe/H] for seven clusters in common with our full sample. For the common seven including NGC 6691 and NGC 7762, the mean difference (Us - C) is a mere $-0.04\pm0.04$. Unfortunately, Fe was the  only element reported on by Casamiquela et al.

\begin{figure*}
\begin{center}
\includegraphics[trim=0.2cm 7.0cm 0.4cm 4.0cm, clip=true,height=0.30\textheight,width=0.70\textheight]{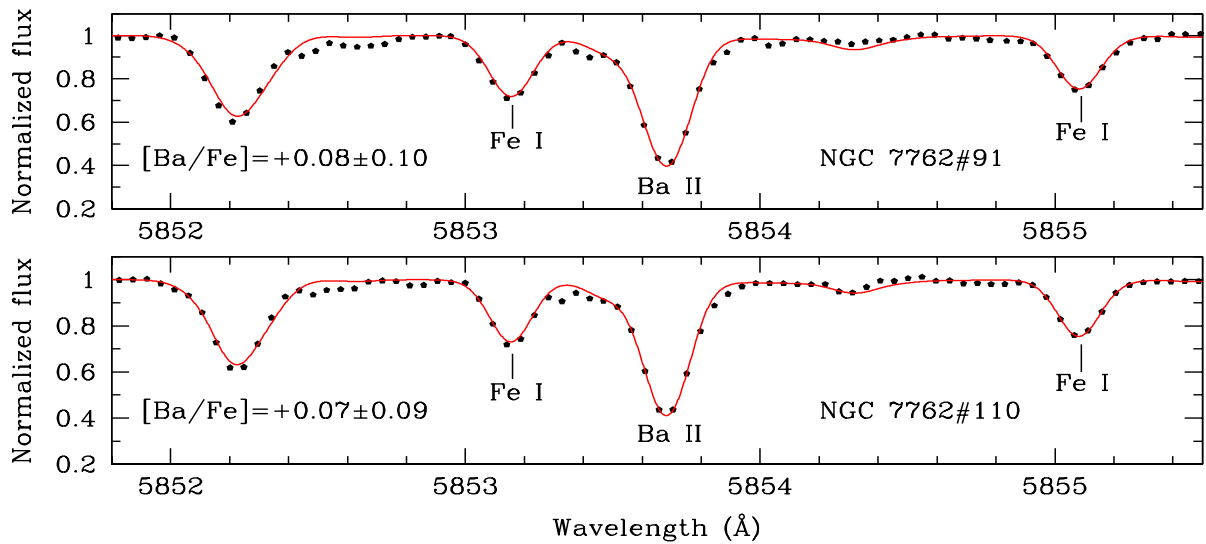}
\caption[]{ Comparison of synthetic spectra (red lines) with the observed spectra (black dots) of stars \#91 and \#110 in the OC NGC 7762 near the Ba {\scs II} lines. In each panel we show the best fit abundance value along with errors due to uncertainties in the best fit value and atmospheric parameters. }
\label{synthba_7762}
\end{center}
\end{figure*}

\subsection{NGC 7762}

The [Fe/H] of NGC 7762 was determined from high-resolution spectra by Casamiquela et al. (2017): they (their Table 9) found [Fe/H] $= 0.01\pm0.04$ from five giants. Our value (Table \ref{mean_abundance}) from three giants is $-0.07\pm0.04$). These three giants are slightly more evolved than the five selected by Casamiquela et al. These [Fe/H] estimates are in good agreement.

An abundance analysis covering Na, Si, Ca, Ti, Fe, Ni and Ba was reported by Carraro, Semenko \& Villanova (2016)  from medium resolution
(R $\simeq 13000$) spectra of the interval 5940--6690\AA. The mean [Fe/H] $= +0.04\pm0.12$ with a range of $-0.15$ to $+0.18$ across the
sample of eight giants. This mean is consistent with our Fe abundance from three giants.

Our sample includes two of the eight giants observed by Carraro et al, namely $\#$91 and $\#$110. Our spectroscopic estimates of the atmospheric
parameters are in good agreement with their work: differences in $T_{\rm eff}$ are within 50 K, surface gravities within 0.1 in $\log g$, and the
microturbulent velocities within 0.1 km s$^{-1}$. Differences in abundances between our and their estimates are 0.07 for [Fe/H], 0.19 for [Na/Fe],
$-0.08$ for [Si/Fe], $-0.12$ for [Ca/Fe], $-0.11$ for [Ti/Fe], and $-0.35$ for [Ba/Fe] for $\#91$. For $\#$110, the differences are $-0.16$ for [Fe/H], 0.23 for [Na/Fe], 0.05 for [Si/Fe], $-0.26$ for [Ca/Fe], $-0.14$ for [Ni/Fe], and $-0.31$ for Ba. These differences appear large and are at odds with the idea that a cluster has a homogeneous composition.

No evidence for a barium overabundance is shown by our spectra. The differences in atmospheric parameters are too small to yield a Ba
abundance difference of 0.3 dex. We provide in Figure \ref{synthba_7762} synthetic spectra fits to the observed spectra of $\#$91 and $\#$110 in the region of the Ba\,{\sc ii} 5854 \AA\ line. The near-solar [Ba/Fe] ratio obtained in our analysis takes into account, as necessary, the hyperfine structure and isotopic shifts in the spectrum synthesis.

\section{Sodium abundances and the first dredge-up}

\subsection{Non-LTE corrections} \label{nonlte_na_al_fe}

Sodium is expected to be enriched in  the atmosphere of a giant thanks to the first dredge-up. Since the enrichment is predicted to be slight, we consider how non-LTE effects may alter our LTE abundances of Na, Al (a useful reference element) and Fe. Then, the non-LTE abundance
ratios [Na/Fe], [Al/Fe] and [Na/Al] are compared with predictions about the first dredge-up.

For all giants in this and earlier papers in this series, the Na abundances derived assuming LTE were corrected for non-LTE effects using the grids of Lind et al. (2011). The corrections were derived on a line-by-line basis, using the stellar parameters (e.g., T$_{\rm eff}$, log~$g$) and the EWs of Na lines (5688.20, 6154.22 and 6160.74 \AA) of each star as input to the interactive non-LTE database\footnote{\url{http://www.inspect-stars.com/}}.

The average non-LTE sodium abundance for all stars including the Sun is lower than its average LTE abundance by values in the range $-$0.09 to $-$0.14 dex. The average non-LTE correction of $-$0.09 dex obtained for the Sun using the above set of three Na lines makes our reference non-LTE Na abundance as log\,$\epsilon$(Na)= 6.26 dex.  As a result, our NLTE Na abundances derived relative to the Sun are not very much different from the differential LTE Na abundances. The non-LTE corrections applied for cluster giants in our analysis reduce the differential LTE Na abundance on an average by 0.0 to $-$0.05 dex.

The non-LTE corrections for Al were computed using the extensive grids of abundance corrections discussed in Nordlander \& Lind (2017). The non-LTE Al abundance calculations were carried out using the stellar parameters and LTE Al abundances for the 7835, 7836, 8773, and 8774 \AA\ lines as input to the source code\footnote{Available online at \url{https://www.mso.anu.edu.au/~thomasn/NLTE/}} of Nordlander \& Lind. We derived an averaged non-LTE Al abundance correction of $-$0.02 dex for the Sun and corrections  in the range $-$0.08 dex to $-$0.17 dex for the cluster giants. Our revised non-LTE Al solar abundance is 6.31 dex. These corrections for the cluster giants lower average LTE Al abundances by values in the range $-$0.06 to $-$0.15 dex.
Table \ref{nlte_naal} lists LTE derived abundances of Na and Al in OCs (this paper and Reddy et al. 2012, 2013, 2015, 2016) and their non-LTE counterparts.

Additionally, we derived non-LTE iron abundances for our sample of giants in the OCs (Lind et al. 2011). At the metallicity of our cluster's giants (0.0 to $-$0.26 dex), we derived a non-LTE Fe\,{\sc i} correction of $+$0.02 dex  and the abundance of 7.54 dex for the Sun and $+$0.02 to $+0.06$ dex for giants using a representative set of iron lines (Fe{\scs I}: the 6240.6, 6252.6, 6498.9, 6574.2, 6609.1, 6739.5, 6750.1, 6793.2; Fe\,{\sc ii}: 5425.3, 6247.6, 6369.5 lines) and for stellar parameters and abundances representative of the giants in our sample of OCs. The non-LTE Fe\, {\sc ii} corrections derived for the Sun and red giants are almost 0.0 dex. The average non-LTE iron abundance corrections derived for cluster giants is about $+$0.03 dex. Therefore, these corrections to our LTE Fe abundance do not affect significantly the global behaviour of NLTE [Na/Fe] with a cluster's turn-off mass.

Not only are non-LTE corrections themselves a source of uncertainty affecting the
Na, Al and Fe abundances  but because 
different non-LTE calculations may have been used in previous papers reporting on Na enrichment comparisons of  published abundances should be adjusted to a common set of non-LTE corrections. Effects of alternative predictions of corrections for non-LTE effects on Na abundances are discussed by Alexeeva, Pakhomov \& Mashonkina (2014) who provide their own calculations for the commonly used Na\,{\sc i} lines. For solar metallicity red giants,  the range in published non-LTE corrections is of the magnitude of the predicted increase from the first dredge-up. Consider  the case of the cluster Collinder 261 where the mass of the red giants is too low for Na enrichment to be expected (see below) but Carretta et al. (2005) reported [Na/Fe] $= +0.33\pm0.06$ after applying non-LTE corrections from Gratton et al. (1999). Smiljanic et al. (2016) remark that this [Na/Fe] becomes [Na/Fe] $\sim 0.0$ after using the more sophisticated calculations  of non-LTE corrections from the Lind et al. (2011) instead of the Gratton et al. grid, a reduction in Na abundance by about 0.3 dex.  (A giant's [Na/Fe] also depends, of course, on the adopted solar Na and Fe abundances and the non-LTE corrections applied to them.)

\begin{figure}
\begin{center}
\includegraphics[trim=0.2cm 4.5cm 12.0cm 4.3cm, clip=true,width=0.47\textwidth,height=0.4\textheight]{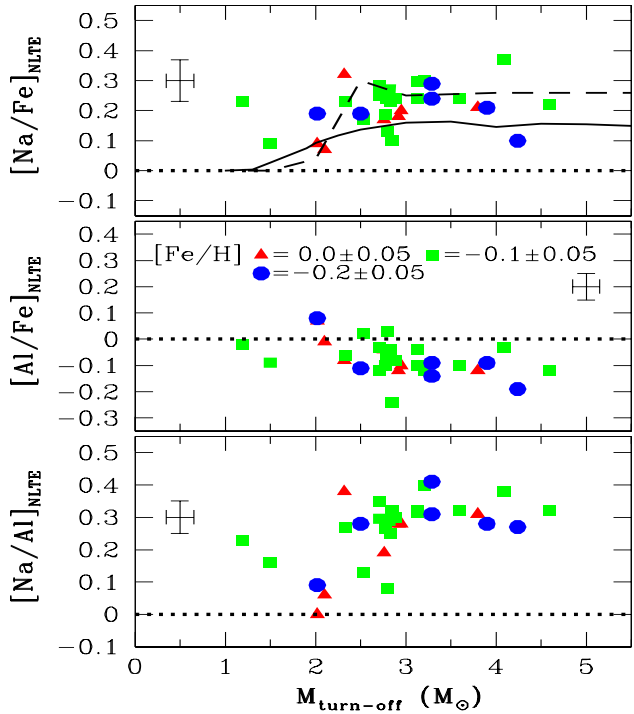}
\caption[]{The non-LTE [Na/Fe] (top panel) and [Al/Fe] (middle panel) and the [Na/Al]-ratio of OCs as a function of cluster's turn-off mass along with a typical error bar as shown (errors of $\pm$0.15 M$_{\rm \odot}$ in stellar mass, $\pm$0.07 dex in [Na/Fe] and $\pm$0.05 dex in [Al/Fe]). In each panel, the red filled triangles, green filled squares and blue filled circles represent, respectively, the clusters with mean [Fe/H] of $0.00\pm0.05$,$-0.10\pm0.05$ and $-0.20\pm0.05$ dex. The continuous solid line (Karakas \& Lattanzio 2014) and the dashed line (Lagarde et al. 2012) in the top panel represent standard model predictions of [Na/Fe] as a function of initial stellar mass (see text for more details). }
\label{na_al_mass}
\end{center}
\end{figure}

\subsection{Sodium enrichment and stellar mass}

The composition of a red clump giant differs from that of its main sequence progenitor thanks to the deep convective envelope
developed as the star ascends the first giant branch. This development results in what is known as the first dredge-up.
Changes in composition primarily affect the elemental and isotopic abundances of C, N and O because these nuclides participate in a series of 
H-burning reactions in the interior regions of the main sequence star subsequently tapped by the convective envelope. Not surprisingly, there is an extensive literature on predicted and observed changes to these nuclidic abundances. Within the span of elements considered here  -- Na to Eu -- sodium is the sole element whose surface abundance is predicted to be detectably increased by the first dredge-up thanks to conversion of $^{22}$Ne to $^{23}$Na  in the interior prior to the first dredge-up. Our initial choice of  predictions for the Na enrichment following the first dredge-up is taken from Karakas \& Lattanzio (2014). The predicted increase in [Na/Fe] is a function of a giant's mass which we identify with the cluster's turn-off mass.

Given that this and our previous papers on open clusters have provided Na abundances for red giants with masses of 1 to 4 $M_\odot$ from a common analysis it seems appropriate to examine whether  Na enrichment in giants follows the theoretical prescription and, particularly, since Karakas \& Lattanzio (2014) in their comprehensive review noted `conflicting results' for Na abundances reported in the literature. Our commentary is not exhaustive because not only do the observations depend on several factors (e.g., the non-LTE corrections to the Na, Al and Fe abundances) but available papers providing predictions concerning  Na enrichments for red giants are few in number and their predictions differ somewhat for reasons not entirely transparent to observers.

In our discussion, we combine Na with Al because  abundances from the Na\,{\sc i} and Al\,{\sc i} lines depend in a quite similar fashion on uncertainties around the atmospheric parameters and all modelling of the first dredge-up predicts that in contrast to Na, the surface abundance of Al is not increased. Thus, the signature of the first dredge-up is an increase of [Na/Fe] but an unchanged [Al/Fe]. Discussion below draws on estimates of the non-LTE abundances of Na, Al and Fe listed in the table \ref{nlte_naal} and are discussed in Section \ref{nonlte_na_al_fe}. Initial comparison of our observed abundances with predictions considers the  mean values for each cluster as a function of a cluster's turn-off mass $M_{\rm turn-off}$ -- see Figure \ref{na_al_mass} where clusters are separated into three categories by their [Fe/H]. Within a single cluster when several giants were analysed, the Na abundances are homogeneous with a typical star-to-star abundance scatter of about 0.05 dex or less. Although the number of giants per cluster is always small, the lack of scatter in [Na/Fe] across a cluster suggests that Na enrichment is not affected greatly by factors which might be expected to vary from star-to-star as, for example, stellar rotation velocity which is predicted to affect the enrichment of Na at the surface.

To begin comparison of  observed and   predicted abundances of Na and Al, a reference set of predictions  for [Na/Fe] is taken from Karakas \& Lattanzio (2014, their Figure 8). All available modelling of the first dredge-up shows that Al is not enriched at the surface. Predictions for [Na/Fe]  shown in Figure \ref{na_al_mass} by the continuous curve refer to stars of metallicity $Z = 0.02$  after the standard first dredge-up. Solar abundances provided by Asplund et al. (2009) correspond to $Z = 0.0142$ which implies that $Z = 0.02$ corresponds to [Fe/H] = $+$0.14, as Karakas \& Lattanzio note, but the predictions are insensitive to $Z$ over the range covered by our clusters. But $Z$ is not the only relevant indicator of a model's composition. Since Na enrichment depends on conversion of $^{22}$Ne to Na, the adopted initial $^{22}$Ne abundance is relevant in predicting the post dredge-up Na abundance. In the solar system, Ne's three stable isotopes have relative abundances of $^{20}$Ne:$^{21}$Ne:$^{22}$Ne of 92.9:0.2:6.8. Karakas \& Lattanzio's $Z = 0.02$ corresponds to a total Ne abundance of 7.97$\pm$0.10 dex on the customary scale where the H abundance is 12.0. In standard models of stellar evolution, surface abundances of red giants evolved beyond the point on the first red giant branch at which the first dredge-up is complete are not
subject to additional change in surface composition until stars evolve up the asymptotic giant branch. Thus, the predictions shown in
Figure \ref{na_al_mass} should apply to our sample of red giants which is dominated by He-core burning  (clump) giants. These predictions for the standard first dredge-up are very similar to those illustrated by Hamdani et al. (2000) for $Z = 0.018$ with, perhaps,  the same nuclear reaction rates adopted by Karakas \& Lattanzio but using a stellar evolution code by Mowlavi (1999) - see also  similar predictions provided by models computed by Ventura et al. (2013) and ilustrated by Smiljanic et al. (2018, Figure 3).\footnote{This statement about Na enrichment following the first dredge-up overlooks the fact that changes of a few percentage points occur in the H and He abundances and these are undetectable at present.  It is assumed that ratios such as [Na/Fe] and [Na/Al] are measurable without serious error even if the model atmosphere assumes slightly inappropriate H and He mass fractions.}

Another set of predictions for He-core burning giants following  the standard first dredge-up is provided by Lagarde et al. (2012, see also Charbonnel \& Lagarde 2010) for an initial composition corresponding to $Z = 0.014$, the solar composition according to Asplund, Grevesse \& Sauval (2005), but with an enhanced Ne abundance (Ne = 8.11 rather than the 2005 value of 7.84) but -- presumably -- the same isotopic fractional abundances. Lagarde et al.'s predictions confirm the onset of Na enrichment at $M_{\rm turn-off} \simeq 2M_\odot$ but yield larger Na enrichments than predicted by Karakas \& Lattanzio at higher masses, i.e., [Na/Fe] $\simeq 0.25$ for $M_{\rm turn-off} \simeq 3M_\odot$ to 6$M_\odot$.  (Smiljanic et al. (2018)'s representation of Lagarde et al.'s predictions for the standard first dredge-up show [Na/Fe] at about 0.27 at 2.5$M_\odot$ increasing to about 0.42 at 6$M_\odot$. Apparently, Smiljanic et al. took these estimates for red giants high up on the AGB following completion of He-core burning.) Predicted [Na/Fe] are not very dependent on the initial metallicity of the stellar models  -- see Smiljanic et al. (2018) who provide predictions for [Fe/H] = 0 and $-0.54$ from Lagarde et al.

The [Na/Fe] predictions from Lagarde et al. appear to provide a closer match to our observations than those from Karakas \& Lattanzio shown in Figure \ref{na_al_mass}. We surmise that the higher [Na/Fe] predicted by Lagarde et al. (2012) are in  part due to their higher adopted Ne abundance. Since Ne is not detectable in the solar photospheric spectrum, its solar abundance is estimated from the solar coronal spectrum or the measurement of solar energetic particles. Lagarde et al. chose to adopt a Ne abundance obtained from spectra of local young B stars where Ne lines are measurable (Cunha, Hubeny \& Lenz 2006).  Predicted [Na/Fe] are also sensitive to the adopted nuclear reaction rates, especially for the $^{22}$Ne(p,$\gamma$)$^{23}$Na reaction.  Hamdani et al. (2000) chose the NACRE rates (Arnould, Goriely \& Jorissen 1999) and find that predicted [Na/Fe] may be substantially changed by adjusting the key rate to its maximum and minimum values, e.g., at $M_{\rm turn-off} = 3M_\odot$, the NACRE rates give [Na/Fe]  = +0.15 but rates adjusted to their maximum and minimum give [Na/Fe] = +0.23 and +0.09, respectively. The $^{22}$Ne(p,$\gamma$) rate has been the subject of extensive recent scrutiny -- see, for example, Ferraro et al. (2018).

The above predictions refer to the standard first dredge-up, i.e., the additional mixing on the first giant branch now commonly referred to as `thermohaline' mixing and mixing in the interior of the main sequence star arising from rapid rotation were not included. Lagarde et al. reported additional calculations in which both thermohaline and rotation-induced mixing were included. A clear indicator that thermohaline  (or an equivalent) mixing occurs in stars is the well known presence of low $^{12}$C/$^{13}$C ratios in
giants in conflict with the higher ratios expected from the standard first dredge-up alone. In contrast to this isotopic ratio, the [Na/Fe] predictions
are little affected  by addition of thermohaline mixing to the physics of the first dredge-up -- see Charbonnel \& Lagarde (2010, Figure 21). Rotation-induced mixing in a main sequence star may raise [Na/Fe] at a given mass and also may lower  the minimum mass for the onset of surface Na enrichment. Charbonnel \& Lagarde (also, their Figure 21) predict that for an initial rotation velocity of 110 km s$^{-1}$ Na enrichment begins at about 1.4$M_\odot$ and post first dredge-up increases [Na/Fe] to about +0.30 from the +0.23  obtained without rotational-induced mixing. In  calculations reported by Lagarde et al. (2012), an initial rotation velocity dependent on initial mass but within the range of 90 to 137 km s$^{-1}$ was adopted. This range is consistent with observed range for low and intermediate mass stars in open clusters. Lagarde et al.'s calculations show that their modelling of thermohaline and rotation-induced mixing raises [Na/Fe] by about 0.06 at 3$M_\odot$ to 0.11 at 6$M_\odot$ for He-core burning giants.

In light of the uncertainties around the predicted Na enrichment from the first dredge-up,
it is fair to conclude that  the predicted [Na/Fe] and [Al/Fe] for giant stars with initial masses from  about 2$M_\odot$ to 5$M_\odot$  match the  observed values for giants from our collection of clusters.  The scatter in Na abundances at a given mass does not exceed the measurement errors and, thus, rotationally-induced mixing is unlikely to be a major contributor inside the main sequence progenitors. An obvious lacuna in our measurements of Na in red giants concerns low mass stars, $M < 2M_\odot$, where Na enrichment is not expected unless rotationally-induced   mixing is severe.
(Two of our clusters have a turn-off mass of less than 1.5$M_\odot$ and one (NGC 2682) appears to be Na enriched beyond expectation. Such a large value of [Na/Fe] measured in our study for NGC 2682 is in agreement with others in the literature (see Table 7 in Reddy et al. 2013).) Exploration of the range below about 1$M_\odot$ is difficult because few clusters survive for the 10 billion years or longer necessary for these low mass stars to become red giants.  Additionally, slowly-acting atomic diffusion can distort the surface compositions of stars at the main sequence turn-off and so corrupt the giant-main sequence comparison.

There are, however, five clusters with $M_{\rm turn-off} \leq 2M_\odot$ for which Na has been measured: Collinder 261,  Trumpler 20, NGC 2243, Berkeley 25 and Ruprecht 147. Collinder 261 was initially (Carretta et al. 2005) reported to be Na rich but Smiljanic et al. (2016) argued that more appropriate non-LTE corrections lead to [Na/Fe] $\sim 0.0$ (see above). Smiljanic et al. (2016) analysed  the three clusters: Trumpler 20 with $M_{\rm turn-off} = 1.8M_\odot$,
NGC 2243 with $M_{\rm turn-off} = 1.2M_\odot$ and Berkeley 25 with $M_{\rm turn-off} = 1.15M_\odot$ obtaining non-LTE [Na/Fe] values of 0.06, 0.10 and 0.04, respectively, and non-LTE [Al/Fe] values of less than 0.07. Ruprecht 147 with $M_{\rm turn-off} \simeq 1.0M_\odot$ was analysed by Bragaglia et al. (2018) obtaining the non-LTE value [Na/Fe] $= 0.24$ for giant stars and 0.08 for main sequence stars or a $+0.16$ increase for the giants. All of these clusters have a metallicity near solar. Unless there are hidden offsets in these analyses, it appears that sodium in giants with masses of less than about 2$M_\odot$ is, as predicted, not enriched.  

Cluster giants with masses above 2$M_\odot$ have been analysed by others. Clusters from the Gaia-ESO survey are discussed by Smiljanic et al. 2016, 2018) with masses $M_{\rm turn-off} = 2.2M_\odot$ to $5.6M_\odot$: the non-LTE [Na/Fe] values are in good agreement with ours with the possible exception of a giant in Trumpler 2 with $M_{\rm turn-off} = 5.6M\odot$ having [Na/Fe] $= 0.48$.  Other clusters with non-LTE [Na/Fe] include  the Hyades (Smiljanic 2012) with $+0.30$ at $M_{\rm turn-off} = 2.5M_\odot$, NGC 4609 (Drazdauskas et al. 2016) with
$+0.33$ at $M_{\rm turn-off} = 5.6M_\odot$, NGC 5316 with $+0.27$ at $M_{\rm turn-off} = 5.0M_\odot$ and IC 4756 (Bagdonas et al. 2018) with $+0.14$ at $M_{\rm turn-off} = 2.2M_\odot$. In general, the non-LTE [Al/Fe] are close to zero.  All of these [Na/Fe] values have been corrected for non-LTE effects using the Lind et al. (2011) grids.

It will be extremely challenging to obtain a precise match to observed Na enrichments because of uncertainties around the theoretical modelling of the first dredge-up in giant stars. Sources of uncertainty include  the unknown initial $^{22}$Ne abundance, the $^{22}$Ne(p,$\gamma$)$^{23}$Na nuclear reaction rate,  the rotationally-induced mixing and the possible presence of atomic diffusion affecting the surface abundances in a cluster's main sequence and turn-off stars which might be considered sources for the initial abundances for the cluster's red giants. On the observational side of the comparison of observation and theory,   key information on surface composition changes resulting from the first dredge-up are provided by measurements of the C, N and O elemental and isotopic abundances. These latter measurements should be integrated with the Na abundances in a comprehensive test of the first dredge-up.  Nonetheless, it is clear that Na but not Al enrichment occurs in red giants about as predicted for stars evolving off the main sequence to the He-core burning stage.

\begin{figure*}
\begin{center}
\includegraphics[trim=0.15cm 11.55cm 1.65cm 4.1cm, clip=true,width=0.98\textwidth,height=0.2\textheight]{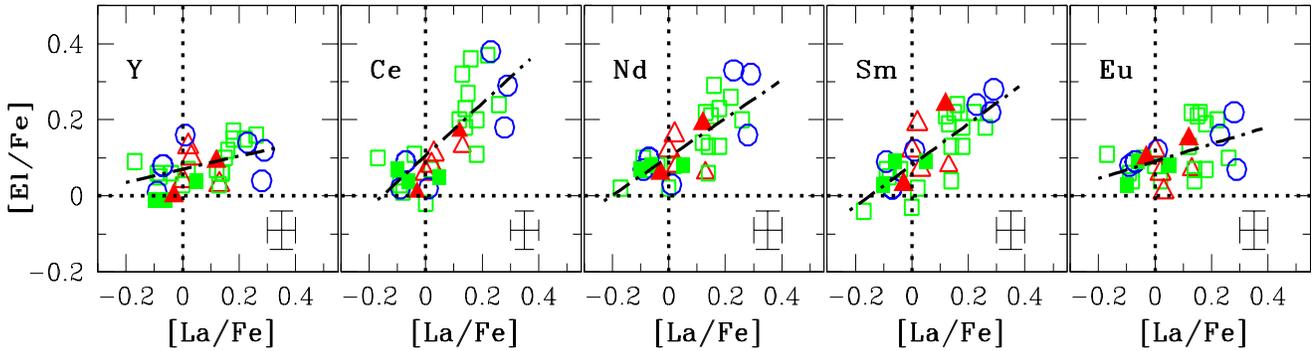}
\caption[]{The relative average cluster abundance ratios [La/Fe] vs [El/Fe] for elements (El) El$=$ Y, Ce, Nd,Sm and Eu. Clusters with mean [Fe/H] of $0.00\pm0.05$ (6 OCs), $-0.10\pm0.05$ (19 OCs) and $-0.20\pm0.05$ (6 OCs) are denoted by red triangles, green squares and blue circles, respectively. The filled and open symbols, respectively, represent the sample of OCs analysed in this paper and those taken from Lambert \& Reddy (2016). The dot-dashed line in each panel has the slope obtained from the least-squares fits with starting and end points set by the abundance data points. }
\label{latoeu_oc}
\end{center}
\end{figure*}

\begin{figure*}
\begin{center}
\includegraphics[trim=0.1cm 10.0cm 4.65cm 4.8cm, clip=true,width=0.9\textwidth,height=0.18\textheight]{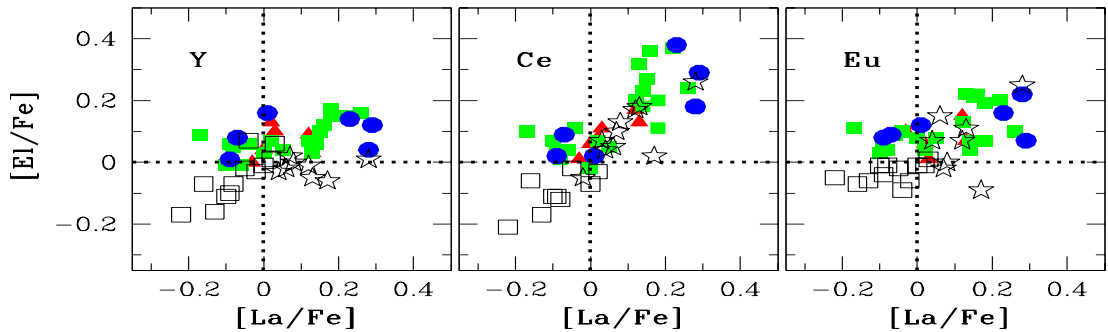}
\caption[]{Same as Figure \ref{latoeu_oc}, but our sample of 31 OCs (5 from this study and 26 from our previous papers) shown as filled symbols are overplotted with Gaia-ESO sample of OCs (Magrini et al. 2018). Gaia-ESO clusters designated as open squares and star symbols, respectively, have the mean [Fe/H] of $=+$0.19$\pm$0.05 dex (12 OCs) and $=-$0.14$\pm$0.12 dex (10 OCs).  }
\label{latoeu_magrini}
\end{center}
\end{figure*}

\begin{figure*}
\begin{center}
\includegraphics[trim=0.1cm 5.3cm 2.65cm 4.8cm, clip=true,width=0.85\textwidth,height=0.35\textheight]{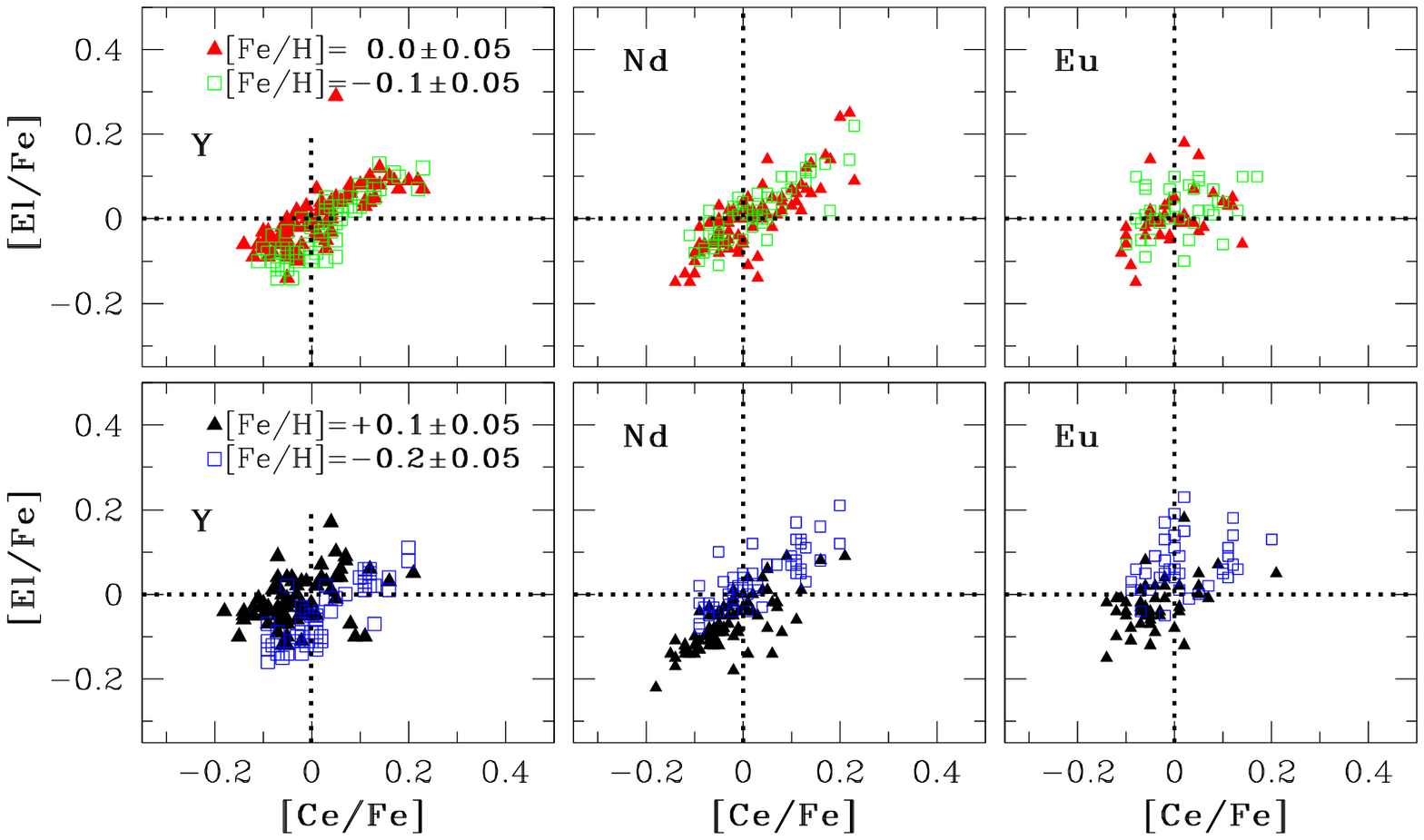}
\caption[]{[Ce/Fe] vs. [El/Fe] for elements (El) El = Y, Nd, and Eu for the field dwarfs from Delgado Mena et al. (2017). Stars with mean [Fe/H] of $0.00\pm0.05$, $-0.10\pm0.05$, $+0.1\pm0.05$, $-$0.20$\pm0.05$ are denoted by red filled triangles, green open squares, black filled triangles, and blue open squares, respectively (see text).  }
\label{latoeu_delgado}
\end{center}
\end{figure*}

\section{Heavy elements as a chemical tag}

Across our previous sample of clusters, the composition of a cluster appeared defined by its [Fe/H] for elements unaffected by the first
dredge-up except for the heavy elements (elements heavier than Fe), i.e., the run of [El/Fe] versus [Fe/H] followed that shown by field dwarfs and giants. The scatter
in [El/Fe] at a given [Fe/H] was small and dominated by the errors of measurement for the samples of cluster giants, field giants and dwarfs. 
This uniformity and small intrinsic scatter in [El/Fe] occurred for elements up through the iron-group was not unexpected because field stars are considered to come from dissolving open clusters whose composition is heavily influenced by contamination by ejecta from supernovae.
The correspondence failed for heavy elements in that elements such as La (for example) whose synthesis occurs in AGB stars showed a range 
in [La/Fe] at a given [Fe/H] with similar ranges found for the giants in clusters and the field. Lambert \& Reddy (2016) speculated that star-forming complexes of interstellar clouds  were polluted by main $s$-process products from AGB stars which in contrast to supernovae eject mass into interstellar clouds at low velocity and
so may contaminate regions of a star-forming complex to varying degrees. Y, a weak $s$-process from massive stars, and Eu, a $r$-process product, do not correlate well with La and the other heavy elements.  This speculative idea received a measure of support from theoretical
considerations by Armillotta, Krumholz \& Fujimoto (2018). Abundances of $s$-process heavy elements appear to provide a chemical tag by which to identify star forming complexes with an otherwise common composition. Unfortunately at the present precision of abundance determinations, the range in heavy element abundances at a given [Fe/H] seems to be continuous and, therefore, tracing field stars to their natal complex is impossible.

Addition of five clusters to the larger sample  considered by Lambert \& Reddy (2016) is unlikely to alter the quantitative evidence about the [El/Fe]
ratios.  Figure \ref{latoeu_oc} shows the key figure from Lambert \& Reddy (their Figure 2) with the five new clusters represented by filled symbols and
previous clusters by open symbols where the choice of a symbol and its color depends on the [Fe/H] of the cluster. Across the total sample of clusters the
[Fe/H] spans the range from $+0.08$ to $-0.44$ but the clusters in Figure \ref{latoeu_oc} cover only the [Fe/H] range from $+0.05$ to $-0.25$ subdivided into three intervals. Figure \ref{latoeu_oc} shows indeed that the new quintet supports the trends previously isolated by Lambert \& Reddy (2016). For giants (and dwarfs) error estimates for the abundances of these heavy elements are essentially identical from Y to Eu because all abundances come from weak lines of singly-charged ions. Thus, the different slopes in the various panels cannot be assigned to a (simple) mischaracterization of a stellar atmosphere.

Lambert \& Reddy (2016) confirmed the trends exhibited in Figure \ref{latoeu_oc} with samples of field dwarfs (Battistini \& Bensby 2016) and of field giants (Mishenina et al. 2006, 2007; Luck 2015) over the same [Fe/H] range  covered by the sample of clusters. Comparisons of
the cluster sample with other samples involving larger intervals of [Fe/H] should recognize two factors: (i) open clusters belong to the
thin disc  and comparisons should not mix thin with thick stars, and (ii) there is evidence that [El/Fe] for heavy elements  is dependent on [Fe/H]
for both dwarfs and giants in the thin disc and, thus, comparisons should compare stars of the same [Fe/H] or correct stars of different [Fe/H] to
their [El/Fe] at a reference [Fe/H].

For dwarfs, abundance determinations for thin disc residents show that [El/Fe] for heavy elements decline slightly with increasing [Fe/H]. Mishenina et al.'s (2016) sample for local dwarfs shows that the decline of [El/Fe] with increasing [Fe/H] for La, Ce, Nd, and Sm, elements dominated by a $s$-process contributions,  has an average slope of about 0.4 dex per dex with a shallower slope for Y.\footnote{Sm is less dominated by the $s$-process than La, Ce and Nd. The main $s$-process provides about 32\% of Sm but higher fractions of La (72\%), Ce (82\%) and Nd (63\%) (Sneden, Cowan \& Gallino 2008; Bisterzo et al. 2014).} Eu, a $r$-process element, has a similar behaviour to Y. The mean slope over the 0.1 dex bins chosen for Figure \ref{latoeu_oc} yields a dispersion of only 0.04 dex but potential samples from the literature may span 0.7 dex in [Fe/H] (see below) and thus involve spreads of 0.3 dex in [El/Fe] which rivals the range shown in Figure \ref{latoeu_oc}. Abundances provided by Battistini \& Bensby (2016) follow the pattern shown by Mishenina et al. - see also Delgado Mena et al. (2017). 

Our first fresh comparison between our results in Figure \ref{latoeu_oc} and others  involves the sample of 22 clusters discussed by Magrini et al. (2018) as part of the extensive Gaia-ESO survey. This sample covers a broader range in [Fe/H] than ours.  The sample is divided at [Fe/H] = 0. For the 12 clusters with [Fe/H] $> 0.0$, the mean [Fe/H] = $+0.19$. For the 10 clusters with [Fe/H] $< 0.0$, the mean [Fe/H] $= -0.14$. Figure \ref{latoeu_magrini} compares our Clusters with those from the Gaia-ESO sample for the common elements Y, La, Ce and Eu. The Gaia-ESO sample with [Fe/H] $< 0.0$ on average should and do fall amongst our `green' and `blue' clusters. The [Fe/H] $> 0$ Gaia-ESO clusters are on average 0.2 dex more [Fe/H]-rich than our `red' clusters. With the standard slope of 0.4 dex per dex, these Gaia-ESO clusters are expected to be displaced from our `red' clusters by about 0.1 tex on average to lower [La/Fe], [Ce/Fe] and [Eu/Fe]. Figure \ref{latoeu_magrini} confirms these shifts. The shift for [Y/Fe] may be somewhat larger than expected. In brief, the Gaia-ESO sample broadly confirms the pattern in Figure \ref{latoeu_oc}.

For the second comparison, we consider the large sample of dwarfs analysed by Delgado Mena et al. (2017). Comparisons in Figure \ref{latoeu_delgado}   are  restricted to stars assigned to the thin disc and in order to minimize abundance errors only to stars with effective temperatures between 5400 K and 6100 K. In this figure Ce is adopted for the abscissa; La was not considered by Delgado Mena et al. We plot Delgado Mena et al.'s results in 0.1 dex
bins with two bins per row. In the top row, the two bins separated by only 0.1 in [Fe/H] overlap with our `red' and `green' clusters. For the standard slope
of $-0.4$ (see above), the scatter within and the displacement between the [Fe/H] $= 0.0$ and $-0.1$ samples should be about 0.04 dex, which is small relative to the observed scatter in the [El/Fe] within the  panels.  In the second row,  the [Fe/H] bins are $+0.1\pm0.05$ and $-0.2\pm0.05$. Again the scatter within a bin attributable to the standard slope of 0.4 dex per dex is only 0.04 which is considerably smaller than the observed scatter. The 0.3 dex difference in [Fe/H] between the two bins should correspond to a displacement of the two distributions by about 0.12  which is seen for Ce, Nd and Eu. Y is more weakly correlated with Ce and Nd.  Eu is almost uncorrelated with Ce. In short, products of the main $s$-process are tightly coupled but products of the weak $s$-process (here, Y) and the $r$-process (here, Eu) appear weakly coupled to the main $s$-process.

\section{Concluding remarks}

Addition of the five open clusters discussed in this paper brings to 33 the total number of clusters whose composition from red giant members has been measured  in this series of papers.  Here, the focus has been on two issues: Na enrichment in red giants and the potential chemical tag offered by heavy element abundances.

First, predictions about Na enrichment of a giant's atmosphere as a result of the first dredge-up have been tested against our measured Na abundances. It is clear that Na is enriched  as predicted but quantitative agreement with predictions has proven elusive very largely because of inherent uncertainties in theoretical predictions about the Na enrichment resulting from the first dredge-up. These uncertainties include a star's initial $^{22}$Ne abundance, the nuclear reaction rate for $^{22}$Ne(p,$\gamma$)$^{23}$Na, the role of rotationally-induced mixing, and the initial Na abundance. Determining or inferring a giant's or its main sequence progenitor's Ne abundance is a challenging problem. For an open cluster, the initial Na abundance may, in principle, be determined from spectra of the cluster's main sequence stars. Such comparison of main sequence stars and giants may not only be vitiated by differential inadequacies of theoretical stellar atmospheres of main sequence and giant stars but also by alteration of the atmospheric composition of main sequence stars by atomic diffusion, as discussed recently for the 4 Gyr cluster M 67 (Bertelli et al. 2018; Gao et al. 2018; Souto et al. 2018). Diffusion is a slow process and is not expected to affect the atmospheric composition of main sequence stars in young open clusters and, indeed, Bertelli et al.'s analysis suggests that effects of diffusion are not present in the 540 Myr cluster NGC 6633.  The first dredge-up by mixing the outer envelope  erases the effects of atomic diffusion.  Future attempts to put the Na enrichment from the first dredge-up on a firm quantitative basis should include determinations of the C, N and O elemental and isotopic abundances. 

Heavy element abundances -- Y, La, Ce, Nd, Sm and Eu -- for the five clusters examined here support our earlier identification that abundances of La, Ce, Nd and Sm with respect to lighter elements (Mg - Ni, for example) may vary by about 0.4 dex (Lambert \& Reddy 2016). Such a variations was found here also among a large sample of FGK dwarfs (Delgado Mena et al. 2017) confirming similar results for field dwarfs and giants provided by Lambert \& Reddy.  In principle, a ratio such as La/Fe may serve as a chemical tag and so identify field stars resulting from a common now-dissolved cluster. This will be very difficult because the run of La etc. abundances for common Mg-Ni abundances appears continuous at the present precision of the abundances. Perhaps, if the abundance data is combined with kinematical information of the precision now provided by Gaia, the chemical-astrometric tag may be able to relate field stars from a common star-forming complex.

There is no reason to cease the exploration of the open cluster population. With the Gaia data  it is possible to make more secure identifications of cluster member including  dwarfs for many additional clusters, even some at larger distances from the Sun than presently explored and, thus, bringing insight into the inner Galaxy and its outer reaches.

\vspace{0.2cm}
\noindent {\bf Acknowledgements:} \\
We thank the referee for thoroughly helpful comments. 
We thank John Lattanzio, Amanda Karakas, Thomas Nordlander and Rodolfo Smiljanic for helpful email exchanges.
We are grateful to the McDonald Observatory's Time Allocation Committee for granting us observing time for this project. This research has made use of the WEBDA database, operated at the Institute for Astronomy of the University of Vienna and the NASA ADS, USA. This research has also made use of Aladin. This publication makes use of data products from the Two Micron All Sky Survey, which is a joint project of the University of Massachusetts and the Infrared Processing and Analysis Center/California Institute of Technology, funded by the National Aeronautics and Space Administration (NASA) and the National Science Foundation (NSF).

\begin{table*}
\caption[Elemental abundances for stars in the OC Stock 2]{The chemical abundances of red giants (ID: 43, 1011, 1082) in the star cluster Stock 2. The abundances measured by synthesis are presented in bold typeface while the remaining elemental abundances were calculated using the line EWs. Numbers in the parentheses indicate the number of lines used in calculating the abundance of that element. } 
\vspace{-0.15cm}
\label{abu_stock2}
\begin{tabular}{lcccc}   \hline
\multicolumn{1}{l}{Species} & \multicolumn{1}{c}{Stock~2$\#$43} & \multicolumn{1}{c}{Stock~2$\#$1011} & \multicolumn{1}{c}{Stock~2$\#$1082} & \multicolumn{1}{c}{Stock~2$_{\mbox{Avg.}}$} \\ \hline

$[$Na I/Fe$]$  &$+0.32\pm0.03$(6)  &$+0.25\pm0.04$(6)   &$+0.25\pm0.02$(5)   &$+0.27\pm0.03$  \\
$[$Mg I/Fe$]$  &$+0.07\pm0.04$(3)  &$\,0.00\pm0.03$(5)  &$+0.01\pm0.01$(4)   &$+0.03\pm0.03$   \\
$[$Al I/Fe$]$  &$+0.07\pm0.03$(5)  &$+0.07\pm0.03$(5)   &$+0.04\pm0.04$(5)   &$+0.06\pm0.03$  \\
$[$Si I/Fe$]$  &$+0.09\pm0.03$(11) &$+0.09\pm0.02$(9)   &$+0.08\pm0.02$(10)  &$+0.09\pm0.02$  \\
$[$Ca I/Fe$]$  &$-0.03\pm0.03$(11) &$-0.04\pm0.03$(9)   &$-0.05\pm0.02$(7)   &$-0.04\pm0.03$  \\
$[$Sc I/Fe$]$  &$+0.02\pm0.03$(5)  &$+0.02\pm0.01$(4)   &$+0.05\pm0.03$(4)   &$+0.03\pm0.02$   \\
$[$Sc II/Fe$]$ &$+0.03\pm0.01$(6)  &$\,0.00\pm0.05$(6)  &$+0.03\pm0.04$(5)   &$+0.02\pm0.04$  \\
$[$Sc II/Fe$]$ &$\bf+0.07\pm0.01$(1)  &$\bf+0.05\pm0.01$(1)   &$\bf+0.07\pm0.01$(1)   &$\bf+0.06\pm0.01$  \\
$[$Ti I/Fe$]$  &$-0.05\pm0.03$(16) &$-0.08\pm0.03$(15)  &$-0.04\pm0.04$(14)  &$-0.06\pm0.03$  \\
$[$Ti II/Fe$]$ &$-0.07\pm0.01$(6)  &$-0.10\pm0.02$(6)   &$-0.07\pm0.03$(6)   &$-0.08\pm0.02$  \\
$[$V I/Fe$]$   &$-0.03\pm0.02$(11) &$-0.04\pm0.03$(12)  &$-0.02\pm0.04$(11)  &$-0.03\pm0.03$  \\
$[$Cr I/Fe$]$  &$+0.01\pm0.04$(8)  &$-0.01\pm0.03$(9)   &$\,0.00\pm0.03$(9)  &$\,0.00\pm0.03$  \\
$[$Cr II/Fe$]$ &$+0.03\pm0.04$(7)  &$+0.05\pm0.02$(7)   &$+0.01\pm0.02$(7)   &$+0.03\pm0.03$  \\
$[$Mn I/Fe$]$  &$\bf-0.06\pm0.01$(2)  &$\bf-0.07\pm0.01$(2)   &$\bf-0.07\pm0.01$(2)   &$\bf-0.07\pm0.01$  \\
$[$Fe I/H$]$   &$-0.07\pm0.03$(157)&$-0.06\pm0.04$(157) &$-0.05\pm0.03$(150) &$-0.06\pm0.03$  \\
$[$Fe II/H$]$  &$-0.07\pm0.02$(20) &$-0.04\pm0.04$(21)  &$-0.06\pm0.02$(18)  &$-0.06\pm0.03$  \\
$[$Co I/Fe$]$  &$-0.01\pm0.02$(6)  &$-0.02\pm0.02$(6)   &$+0.01\pm0.01$(6)   &$-0.01\pm0.02$  \\
$[$Ni I/Fe$]$  &$-0.02\pm0.04$(22) &$-0.03\pm0.04$(23)  &$-0.03\pm0.04$(22)  &$-0.03\pm0.04$  \\
$[$Cu I/Fe$]$  &$\bf-0.11\pm0.01$(1)  &$\bf-0.12\pm0.01$(1)   &$\bf-0.10\pm0.01$(1)   &$\bf-0.11\pm0.01$  \\
$[$Zn I/Fe$]$  &$\bf-0.02\pm0.01$(1)  &$\bf-0.04\pm0.01$(1)   &$\bf-0.03\pm0.01$(1)   &$\bf-0.03\pm0.01$  \\
$[$Y II/Fe$]$  &$\,0.00\pm0.03$(10)&$-0.02\pm0.02$(10)  &$-0.02\pm0.02$(8)   &$-0.01\pm0.02$  \\
$[$Zr I/Fe$]$  & $+0.02\pm0.04$(6) &$+0.03\pm0.04$(6)   &$+0.04\pm0.02$(5)   &$+0.03\pm0.03$  \\
$[$Zr II/Fe$]$ & $+0.05\pm0.04$(3) &$+0.04\pm0.01$(3)   &$\,0.00\pm0.02$(3)  &$+0.03\pm0.03$  \\
$[$Ba II/Fe$]$ &$\bf+0.27\pm0.02$(1)  &$\bf+0.25\pm0.02$(1)   &$\bf+0.26\pm0.02$(1)   &$\bf+0.26\pm0.02$  \\
$[$La II/Fe$]$ & $-0.07\pm0.01$(5) &$-0.06\pm0.04$(5)   &$-0.06\pm0.01$(5)   &$-0.06\pm0.02$  \\
$[$Ce II/Fe$]$ & $+0.02\pm0.02$(5) &$+0.04\pm0.04$(5)   &$+0.05\pm0.02$(5)   &$+0.04\pm0.03$  \\
$[$Nd II/Fe$]$ & $+0.06\pm0.04$(10)&$+0.09\pm0.02$(10)  &$+0.10\pm0.04$(9)   &$+0.08\pm0.03$  \\
$[$Sm II/Fe$]$ & $+0.09\pm0.01$(7) &$+0.09\pm0.02$(7)   &$+0.10\pm0.03$(7)   &$+0.09\pm0.02$  \\
$[$Eu II/Fe$]$ &$\bf+0.07\pm0.01$(1)  &$\bf+0.09\pm0.01$(1)   &$\bf+0.10\pm0.01$(1)   &$+0.09\pm0.01$  \\

\hline
\end{tabular}
\end{table*}

\begin{table*}
\caption[Elemental abundances for stars in the OC NGC 2168]{Same as Table \ref{abu_stock2} but for chemical abundances of stars (IDs: 81, 310, and 662) in NGC 2168 and stars (ID: 134 and HD 162496) in NGC 6475. } 
\label{abu_NGC2168_6475}
{\fontsize{8}{10}\selectfont
\begin{tabular}{llllllll}   \hline
\multicolumn{1}{l}{Species} & \multicolumn{1}{l}{NGC~2168$\#$81} & \multicolumn{1}{l}{NGC~2168$\#$310} & \multicolumn{1}{l}{NGC~2168$\#$662} 
& \multicolumn{1}{l}{NGC~2168$_{\mbox{Avg.}}$} \vline  & \multicolumn{1}{l}{ HD~162391} & \multicolumn{1}{l}{ HD~162496} & \multicolumn{1}{l}{NGC~6475$_{\mbox{Avg.}}$}   \\   \hline

$[$Na I/Fe$]$ &$+0.20\pm0.02$(6) &$+0.27\pm0.03$(4) &$+0.18\pm0.03$(6) &$+0.22\pm0.03$ &$+0.23\pm0.02$(6) &$+0.22\pm0.03$(6) &$+0.23\pm0.02$  \\
$[$Mg I/Fe$]$ &$+0.05\pm0.03$(5) &$+0.05\pm0.04$(5) &$+0.08\pm0.03$(5) &$+0.06\pm0.03$ &$+0.01\pm0.03$(5) &$\,0.00\pm0.03$(5) &$+0.01\pm0.03$   \\
$[$Al I/Fe$]$ &$+0.02\pm0.02$(5) &$+0.06\pm0.03$(5) &$+0.01\pm0.03$(5) &$+0.03\pm0.03$ &$+0.04\pm0.02$(5) &$+0.03\pm0.02$(5) &$+0.04\pm0.02$  \\
$[$Si I/Fe$]$ &$+0.33\pm0.04$(9) &$+0.24\pm0.02$(12) &$+0.29\pm0.04$(9) &$+0.28\pm0.03$ &$+0.05\pm0.03$(11) &$+0.06\pm0.04$(9) &$+0.06\pm0.03$  \\
$[$Ca I/Fe$]$ &$+0.01\pm0.04$(11) &$+0.03\pm0.04$(11) &$\,0.00\pm0.02$(10) &$+0.01\pm0.03$ &$-0.03\pm0.03$(12) &$-0.02\pm0.03$(10) &$-0.03\pm0.03$  \\
$[$Sc I/Fe$]$ &   $\ldots$      &    $\ldots$     &    $\ldots$        &    $\ldots$     &$+0.03\pm0.03$(3) &$-0.02\pm0.03$(3) &$+0.00\pm0.03$   \\
$[$Sc II/Fe$]$ &$-0.04\pm0.03$(6) &$-0.08\pm0.03$(4) &$-0.01\pm0.03$(5) &$-0.04\pm0.03$ &$-0.03\pm0.02$(5) &$\,0.00\pm0.02$(6) &$-0.01\pm0.02$  \\
$[$Sc II/Fe$]$ &$\bf-0.04\pm0.01$(1) &$\bf-0.03\pm0.01$(1) &$\bf-0.05\pm0.01$(1) &$\bf-0.04\pm0.01$ &$\bf-0.02\pm0.02$(1) &$\bf-0.05\pm0.02$(1) &$\bf-0.04\pm0.02$  \\
$[$Ti I/Fe$]$ &$-0.13\pm0.03$(14) &$-0.06\pm0.02$(13) &$-0.16\pm0.03$(13) &$-0.12\pm0.03$ &$\,0.00\pm0.03$(13)&$-0.04\pm0.03$(14) &$-0.02\pm0.03$  \\
$[$Ti II/Fe$]$ &$-0.12\pm0.01$(7) &$-0.10\pm0.03$(6) &$-0.10\pm0.03$(7) &$-0.11\pm0.02$ &$-0.04\pm0.02$(6) &$-0.05\pm0.02$(5) &$-0.04\pm0.02$  \\
$[$V I/Fe$]$ &$-0.05\pm0.04$(11) &$-0.03\pm0.04$(8) &$-0.07\pm0.04$(11) &$-0.05\pm0.04$ &$\,0.00\pm0.04$(8) &$-0.01\pm0.03$(11) &$-0.01\pm0.03$  \\
$[$Cr I/Fe$]$ &$-0.03\pm0.02$(8) &$\,0.00\pm0.03$(7) &$-0.03\pm0.03$(8) &$-0.02\pm0.03$ &$\,0.00\pm0.03$(8) &$+0.01\pm0.03$(8) &$+0.01\pm0.03$  \\
$[$Cr II/Fe$]$ &$-0.01\pm0.03$(7) &$\,0.00\pm0.03$(4) &$+0.02\pm0.03$(6) &$\,0.00\pm0.03$ &$+0.01\pm0.03$(7) &$+0.03\pm0.02$(6) &$+0.02\pm0.02$  \\
$[$Mn I/Fe$]$ &$\bf-0.04\pm0.02$(2) &$\bf-0.05\pm0.02$(2) &$\bf-0.10\pm0.02$(2) &$\bf-0.06\pm0.02$ &$\bf-0.04\pm0.01$(2) &$\bf-0.08\pm0.02$(2) &$\bf-0.06\pm0.01$  \\
$[$Fe I/H$]$ &$-0.12\pm0.03$(150)&$-0.09\pm0.03$(146) &$-0.11\pm0.03$(153) &$-0.11\pm0.03$ &$-0.01\pm0.03$(154)&$\,0.00\pm0.03$(153)&$-0.01\pm0.03$  \\
$[$Fe II/H$]$ &$-0.11\pm0.03$(17) &$-0.11\pm0.03$(11) &$-0.10\pm0.02$(17) &$-0.11\pm0.03$ &$-0.01\pm0.03$(16) &$-0.01\pm0.04$(18) &$-0.01\pm0.03$  \\
$[$Co I/Fe$]$ &$+0.02\pm0.02$(5) &$\,0.00\pm0.01$(4) &$+0.06\pm0.02$(5) &$+0.03\pm0.02$ &$+0.01\pm0.02$(5) &$+0.01\pm0.03$(6) &$+0.01\pm0.02$  \\
$[$Ni I/Fe$]$ &$-0.02\pm0.04$(21) &$-0.04\pm0.03$(22) &$-0.01\pm0.03$(21) &$-0.02\pm0.03$ &$-0.03\pm0.03$(24) &$-0.05\pm0.04$(21) &$-0.04\pm0.03$  \\
$[$Cu I/Fe$]$ &$\bf-0.09\pm0.01$(1) &$\bf-0.15\pm0.02$(1) &$\bf-0.10\pm0.01$(1) &$\bf-0.11\pm0.01$ &$\bf-0.19\pm0.02$(1) &$\bf-0.15\pm0.02$(1) &$\bf-0.17\pm0.02$  \\
$[$Zn I/Fe$]$ &$\bf-0.14\pm0.01$(1) &$\bf-0.12\pm0.02$(1) &$\bf-0.15\pm0.02$(1) &$\bf-0.14\pm0.02$ &$\bf-0.19\pm0.02$(1) &$\bf-0.25\pm0.02$(1) &$\bf-0.22\pm0.02$  \\
$[$Y II/Fe$]$ &$\,0.00\pm0.03$(9) &$-0.03\pm0.00$(7) &$\,0.00\pm0.03$(10) &$-0.01\pm0.02$ &$+0.01\pm0.03$(8) &$+0.01\pm0.02$(9) &$\,0.00\pm0.02$  \\
$[$Zr I/Fe$]$ & $-0.09\pm0.03$(6) &$-0.12\pm0.02$(2) &$-0.12\pm0.04$(4) &$-0.11\pm0.03$ & $+0.09\pm0.02$(6) &$-0.01\pm0.03$(6) &$+0.04\pm0.02$  \\
$[$Zr II/Fe$]$ & $+0.20\pm0.04$(3) &$+0.17\pm0.04$(3) &$+0.16\pm0.02$(3) &$+0.18\pm0.03$ & $+0.06\pm0.03$(3) &$+0.08\pm0.03$(3) &$+0.07\pm0.03$  \\
$[$Ba II/Fe$]$ &$\bf+0.20\pm0.03$(1) &$\bf+0.60\pm0.04$(1) &$\bf+0.15\pm0.02$(1) &$\bf+0.17\pm0.03$ &$\bf+0.11\pm0.02$(1) &$\bf+0.12\pm0.02$(1) &$\bf+0.12\pm0.02$  \\
$[$La II/Fe$]$ & $-0.11\pm0.04$(5) &$-0.10\pm0.01$(5) &$-0.08\pm0.03$(5) &$-0.10\pm0.03$ & $-0.03\pm0.04$(5) &$-0.03\pm0.03$(4) &$-0.03\pm0.03$  \\
$[$Ce II/Fe$]$ & $+0.08\pm0.02$(5) &$+0.06\pm0.02$(4) &$+0.08\pm0.02$(5) &$+0.07\pm0.02$ & $-0.01\pm0.03$(5) &$+0.02\pm0.03$(5) &$+0.01\pm0.03$  \\
$[$Nd II/Fe$]$ & $+0.08\pm0.02$(9) &$+0.05\pm0.01$(8) &$+0.09\pm0.03$(8) &$+0.07\pm0.02$ & $+0.05\pm0.03$(8) &$+0.07\pm0.03$(9) &$+0.06\pm0.03$  \\
$[$Sm II/Fe$]$ & $+0.04\pm0.02$(6) &$+0.02\pm0.03$(7) &$+0.02\pm0.01$(7) &$+0.03\pm0.02$ & $+0.03\pm0.03$(7) &$+0.02\pm0.03$(7) &$+0.03\pm0.03$  \\
$[$Eu II/Fe$]$ &$\bf+0.06\pm0.02$(1) &$\bf\,0.00\pm0.02$(1) &$\bf+0.04\pm0.02$(1) &$\bf+0.03\pm0.02$ &$\bf+0.10\pm0.02$(1) &$\bf+0.11\pm0.02$(1) &$\bf+0.10\pm0.02$  \\

\hline
\end{tabular}
 }
\flushleft{ {\bf Note:} We excluded the barium abundance of the star $\#$310 in NGC 2168 in calculating the cluster mean [Ba/Fe] value due to star's rapid rotation (V$_{\rm rot}$$\sim$\,12 km s$^{-1}$) that may lead to turbulent superficial stellar layers from where the Ba {\scs II} lines emerge. As a result, the inclusion of microturbulence derived from Fe lines in the spectrum synthesis of Ba {\scs II} line is insufficient to represent the true line broadening imposed by relatively turbulent upper photospheric layers on the Ba {\scs II} line which may result in a serious overestimation of Ba abundance by standard LTE abundance analysis (Reddy \& Lambert 2017).}
\end{table*}

\begin{table*}
\caption[Elemental abundances for stars in the OC NGC 6991]{Same as Table \ref{abu_stock2} but for chemical abundances of stars (IDs: 22, 67, 100 and 131) in NGC 6991. } 
\vspace{-0.1cm}
\label{abu_6991}
\begin{tabular}{lccccc}   \hline
\multicolumn{1}{l}{Species} & \multicolumn{1}{c}{NGC~6991$\#$22} & \multicolumn{1}{c}{NGC~6991$\#$67} & \multicolumn{1}{c}{NGC~6991$\#$100} & \multicolumn{1}{c}{NGC~6991$\#$131} & \multicolumn{1}{c}{NGC~6991$_{\mbox{Avg.}}$}  \\ \hline

$[$Na I/Fe$]$  &$+0.06\pm0.02$(6)  &$+0.10\pm0.03$(6)   &$+0.12\pm0.03$(6)   &$+0.11\pm0.02$(6)  &$+0.10\pm0.02$ \\
$[$Mg I/Fe$]$  &$-0.07\pm0.03$(5)  &$-0.06\pm0.03$(7)   &$-0.06\pm0.03$(7)   &$-0.07\pm0.03$(7)  &$-0.06\pm0.03$  \\
$[$Al I/Fe$]$  &$-0.11\pm0.04$(5)  &$-0.06\pm0.02$(5)   &$-0.07\pm0.02$(5)   &$-0.09\pm0.02$(5)  &$+0.08\pm0.03$ \\
$[$Si I/Fe$]$  &$+0.01\pm0.04$(10) &$+0.04\pm0.03$(12)  &$+0.06\pm0.03$(14)  &$+0.06\pm0.03$(14) &$+0.04\pm0.03$ \\
$[$Ca I/Fe$]$  &$\,0.00\pm0.04$(12)&$~0.00\pm0.03$(11)  &$+0.04\pm0.03$(12)  &$+0.02\pm0.02$(12) &$+0.01\pm0.03$ \\
$[$Sc I/Fe$]$  &$\,0.00\pm0.02$(5) &$+0.03\pm0.03$(4)   &$\,0.00\pm0.03$(6)  &$+0.01\pm0.03$(4)  &$+0.01\pm0.03$ \\
$[$Sc II/Fe$]$ &$-0.01\pm0.03$(6)  &$-0.02\pm0.03$(6)   &$-0.02\pm0.02$(5)   &$-0.02\pm0.04$(7)  &$-0.02\pm0.03$ \\
$[$Sc II/Fe$]$ &$\bf+0.01\pm0.01$(1)  &$\bf+0.02\pm0.01$(1)   &$\bf+0.00\pm0.01$(1)   &$\bf+0.02\pm0.01$(1)  &$\bf+0.01\pm0.01$ \\
$[$Ti I/Fe$]$  &$-0.02\pm0.02$(12) &$-0.01\pm0.03$(14)  &$-0.03\pm0.02$(14)  &$\,0.00\pm0.03$(15)&$-0.01\pm0.02$ \\
$[$Ti II/Fe$]$ &$-0.07\pm0.01$(10) &$-0.04\pm0.03$(10)  &$-0.08\pm0.04$(11)  &$-0.06\pm0.02$(11) &$-0.06\pm0.03$ \\
$[$V I/Fe$]$   &$-0.04\pm0.03$(11) &$-0.02\pm0.03$(11)  &$-0.03\pm0.03$(12)  &$-0.01\pm0.03$(13) &$-0.02\pm0.03$ \\
$[$Cr I/Fe$]$  &$-0.03\pm0.02$(8)  &$-0.01\pm0.03$(10)  &$-0.01\pm0.02$(9)   &$-0.03\pm0.02$(8)  &$-0.02\pm0.02$ \\
$[$Cr II/Fe$]$ &$-0.03\pm0.02$(7)  &$+0.03\pm0.02$(7)   &$+0.04\pm0.02$(8)   &$+0.05\pm0.02$(8)  &$+0.02\pm0.02$ \\
$[$Mn I/Fe$]$  &$\bf-0.01\pm0.01$(2)  &$\bf+0.00\pm0.01$(2)   &$\bf+0.00\pm0.01$(2)   &$\bf-0.01\pm0.01$(2)  &$\bf-0.01\pm0.01$ \\
$[$Fe I/H$]$   &$+0.01\pm0.04$(143)&$-0.01\pm0.03$(162) &$+0.01\pm0.03$(159) &$+0.01\pm0.02$(160)&$+0.01\pm0.03$ \\
$[$Fe II/H$]$  &$+0.01\pm0.03$(15) &$~0.00\pm0.03$(20)  &$-0.01\pm0.02$(20)  &$+0.01\pm0.03$(22) &$\,0.00\pm0.03$ \\
$[$Co I/Fe$]$  &$-0.02\pm0.03$(6)  &$+0.02\pm0.02$(6)   &$+0.01\pm0.03$(6)   &$+0.01\pm0.02$(6)  &$+0.01\pm0.02$ \\
$[$Ni I/Fe$]$  &$-0.01\pm0.02$(22) &$-0.01\pm0.02$(23)  &$\,0.00\pm0.03$(23) &$-0.01\pm0.03$(23) &$-0.01\pm0.02$ \\
$[$Cu I/Fe$]$  &$\bf-0.05\pm0.01$(1)  &$\bf-0.03\pm0.01$(1)   &$\bf-0.02\pm0.01$(1)   &$\bf-0.03\pm0.01$(1)  &$\bf-0.03\pm0.01$ \\
$[$Zn I/Fe$]$  &$\bf\,0.00\pm0.01$(1) &$\bf+0.01\pm0.01$(1)   &$\bf+0.02\pm0.01$(1)   &$\bf+0.01\pm0.01$(1)  &$\bf+0.01\pm0.01$ \\
$[$Y II/Fe$]$  &$+0.08\pm0.03$(9) &$+0.09\pm0.03$(10)  &$+0.08\pm0.03$(10)  &$+0.09\pm0.02$(10) &$+0.08\pm0.03$ \\
$[$Zr I/Fe$]$  &$+0.12\pm0.04$(5) &$+0.12\pm0.02$(6)   &$+0.09\pm0.03$(5)   &$+0.12\pm0.03$(5)  &$+0.011\pm0.03$ \\
$[$Zr II/Fe$]$ &$+0.08\pm0.03$(3) &$+0.10\pm0.01$(3)   &$+0.07\pm0.02$(3)   &$+0.11\pm0.03$(3)  &$+0.09\pm0.02$ \\
$[$Ba II/Fe$]$ &$\bf+0.12\pm0.03$(1)  &$\bf+0.13\pm0.03$(1)   &$\bf+0.12\pm0.02$(1)   &$\bf+0.12\pm0.02$(1)  &$\bf+0.12\pm0.02$ \\
$[$La II/Fe$]$ &$+0.10\pm0.02$(5) &$+0.13\pm0.04$(7)   &$+0.14\pm0.03$(6)   &$+0.13\pm0.01$(6)  &$+0.12\pm0.03$ \\
$[$Ce II/Fe$]$ &$+0.17\pm0.03$(5) &$+0.17\pm0.03$(5)   &$+0.17\pm0.03$(5)   &$+0.16\pm0.03$(5)  &$+0.17\pm0.03$ \\
$[$Nd II/Fe$]$ &$+0.20\pm0.03$(6) &$+0.20\pm0.02$(10)  &$+0.18\pm0.03$(11)  &$+0.17\pm0.03$(10) &$+0.19\pm0.03$ \\
$[$Sm II/Fe$]$ &$+0.22\pm0.04$(6) &$+0.26\pm0.04$(7)   &$+0.23\pm0.03$(7)   &$+0.23\pm0.03$(7)  &$+0.24\pm0.03$ \\
$[$Eu II/Fe$]$ &    $\ldots$      &$\bf+0.16\pm0.02$(1) &$\bf+0.15\pm0.02$(1) &$\bf+0.14\pm0.02$(1) &$\bf+0.15\pm0.02$ \\

\hline
\end{tabular}
\end{table*}


\begin{table*}
\caption[Elemental abundances for stars in the OC NGC 7762]{Same as Table \ref{abu_stock2} but for chemical abundances of stars (IDs: 22, 67, 100 and 131) in NGC 6991. } 
\vspace{-0.1cm}
\label{abu_ngc7762}
\begin{tabular}{lcccc}   \hline
\multicolumn{1}{l}{Species} & \multicolumn{1}{c}{NGC~7762$\#$35} & \multicolumn{1}{c}{NGC~7762$\#$91} & \multicolumn{1}{c}{NGC~7762$\#$110} & \multicolumn{1}{c}{NGC~7762$_{\mbox{Avg.}}$}  \\ \hline

$[$Na I/Fe$]$  &$+0.03\pm0.02$(4)  &$+0.22\pm0.03$(4)   &$+0.09\pm0.02$(4)   &$+0.11\pm0.02$  \\
$[$Mg I/Fe$]$  &$+0.02\pm0.01$(3)  &$+0.04\pm0.02$(4)   &$+0.04\pm0.03$(4)   &$+0.03\pm0.02$   \\
$[$Al I/Fe$]$  &$+0.06\pm0.01$(5)  &$+0.01\pm0.04$(5)   &$+0.07\pm0.02$(4)   &$+0.05\pm0.03$  \\
$[$Si I/Fe$]$  &$+0.28\pm0.04$(13) &$+0.21\pm0.04$(11)  &$+0.20\pm0.02$(10)  &$+0.23\pm0.03$  \\
$[$Ca I/Fe$]$  &$-0.04\pm0.03$(12) &$-0.01\pm0.03$(11)  &$\,0.00\pm0.03$(9)  &$-0.02\pm0.03$  \\
$[$Sc II/Fe$]$ &$-0.02\pm0.04$(7)  &$+0.02\pm0.04$(6)   &$+0.01\pm0.04$(6)   &$\,0.00\pm0.04$  \\
$[$Sc II/Fe$]$ &$\bf+0.06\pm0.01$(1)  &$\bf+0.08\pm0.01$(1)  &$\bf+0.07\pm0.01$(1) &$\bf+0.07\pm0.01$  \\
$[$Ti I/Fe$]$  &$-0.06\pm0.03$(9)  &$-0.09\pm0.03$(13)  &$-0.03\pm0.02$(12)  &$-0.06\pm0.03$  \\
$[$Ti II/Fe$]$ &$-0.07\pm0.03$(8)  &$-0.05\pm0.03$(8)   &$-0.05\pm0.04$(7)   &$-0.06\pm0.03$  \\
$[$V I/Fe$]$   &$\,0.00\pm0.04$(7) &$\,0.00\pm0.03$(12) &$+0.03\pm0.04$(8)   &$+0.01\pm0.04$  \\
$[$Cr I/Fe$]$  &$-0.01\pm0.03$(10) &$\,0.00\pm0.02$(8)  &$+0.03\pm0.03$(10)  &$+0.01\pm0.03$  \\
$[$Cr II/Fe$]$ &$+0.05\pm0.04$(6)  &$+0.08\pm0.03$(8)   &$+0.07\pm0.04$(6)   &$+0.07\pm0.04$  \\
$[$Mn I/Fe$]$  &$\bf-0.14\pm0.01$(2)  &$\bf-0.11\pm0.01$(2) &$\bf-0.13\pm0.01$(2)  &$\bf-0.13\pm0.01$  \\
$[$Fe I/H$]$   &$-0.06\pm0.03$(141)&$-0.08\pm0.03$(140) &$-0.07\pm0.04$(133) &$-0.07\pm0.03$  \\
$[$Fe II/H$]$  &$-0.07\pm0.02$(17) &$-0.08\pm0.03$(16)  &$-0.08\pm0.04$(15)  &$-0.08\pm0.03$  \\
$[$Co I/Fe$]$  &$+0.12\pm0.01$(5)  &$+0.11\pm0.03$(6)   &$+0.11\pm0.03$(6)   &$+0.11\pm0.02$  \\
$[$Ni I/Fe$]$  &$+0.01\pm0.03$(23) &$-0.01\pm0.04$(19)  &$+0.03\pm0.03$(17)  &$+0.01\pm0.03$  \\
$[$Cu I/Fe$]$  &$\bf+0.02\pm0.01$(1)  &$\bf+0.03\pm0.01$(1)  &$\bf+0.04\pm0.01$(1)  &$\bf+0.03\pm0.01$  \\
$[$Zn I/Fe$]$  &$\bf-0.09\pm0.01$(1)  &$\bf-0.06\pm0.01$(1)  &$\bf-0.06\pm0.01$(1)  &$\bf-0.07\pm0.01$  \\
$[$Y II/Fe$]$  &$+0.04\pm0.03$(7) &$+0.06\pm0.03$(9)   &$+0.03\pm0.03$(9)   &$+0.04\pm0.03$  \\
$[$Zr I/Fe$]$  &$-0.02\pm0.04$(6) &$-0.02\pm0.02$(6)   &$+0.02\pm0.02$(6)   &$+0.01\pm0.03$  \\
$[$Zr II/Fe$]$ &$-0.01\pm0.02$(3) &$+0.03\pm0.01$(3)   &$+0.07\pm0.03$(3)   &$+0.03\pm0.02$  \\
$[$Ba II/Fe$]$ &$\bf+0.06\pm0.02$(1)  &$\bf+0.08\pm0.02$(1)  &$\bf+0.07\pm0.02$(1) &$\bf+0.07\pm0.02$  \\
$[$La II/Fe$]$ &$+0.05\pm0.04$(7) &$+0.05\pm0.03$(6)   &$+0.04\pm0.03$(5)   &$+0.05\pm0.03$  \\
$[$Ce II/Fe$]$ &$+0.05\pm0.04$(5) &$+0.04\pm0.02$(4)   &$+0.06\pm0.02$(5)   &$+0.05\pm0.03$  \\
$[$Nd II/Fe$]$ &$+0.07\pm0.03$(9) &$+0.09\pm0.03$(7)   &$+0.09\pm0.03$(9)   &$+0.08\pm0.03$  \\
$[$Sm II/Fe$]$ &$+0.08\pm0.03$(7) &$+0.10\pm0.04$(7)   &$+0.10\pm0.04$(7)   &$+0.09\pm0.04$  \\
$[$Eu II/Fe$]$ &$\bf+0.06\pm0.01$(1) &$\bf+0.10\pm0.01$(1) &$\bf+0.10\pm0.01$(1) &$\bf+0.08\pm0.01$  \\

\hline
\end{tabular}
\end{table*}

\begin{table*}
\caption[Non-LTE Na and Al abundances]{Mean LTE and non-LTE abundances of Na and Al in OCs analysed in this paper and in Reddy et al. (2012, 2013, 2015, 2016).} 
\vspace{-0.15cm}
\label{nlte_naal}
\begin{tabular}{lccccc}   \hline
\multicolumn{1}{l}{Cluster} & \multicolumn{1}{c}{[Na/Fe]} & \multicolumn{1}{c}{[Na/Fe]} & \multicolumn{1}{c}{[Al/Fe]} & \multicolumn{1}{c}{[Al/Fe]} & \multicolumn{1}{c}{Mass$_{turn-off}$}  \\ 
\multicolumn{1}{l}{ } & \multicolumn{1}{c}{(LTE)} & \multicolumn{1}{c}{(non-LTE)} & \multicolumn{1}{c}{(LTE)} & \multicolumn{1}{c}{(non-LTE)} & \multicolumn{1}{c}{(M$_\odot$)}  \\
\hline
 
\multicolumn{6}{c}{[Fe/H]$=$0.00$\pm$0.05 dex}           \\
 NGC 6991 &   0.10$\pm$0.01 &   0.06$\pm$0.01 &   0.08$\pm$0.03 &   0.00$\pm$0.03 &   2.10  \\
 NGC 6475 &   0.23$\pm$0.02 &   0.20$\pm$0.02 &   0.04$\pm$0.03 &  -0.11$\pm$0.03 &   3.80  \\
 NGC 2099 &   0.23$\pm$0.02 &   0.19$\pm$0.02 &   0.01$\pm$0.03 &  -0.09$\pm$0.03 &   2.95  \\
 NGC  752 &   0.12$\pm$0.01 &   0.08$\pm$0.01 &   0.15$\pm$0.02 &   0.08$\pm$0.02 &   2.02  \\
 NGC 6633 &   0.20$\pm$0.01 &   0.16$\pm$0.01 &   0.06$\pm$0.02 &  -0.03$\pm$0.02 &   2.76  \\
 NGC 2281 &   0.20$\pm$0.02 &   0.17$\pm$0.02 &  -0.02$\pm$0.02 &  -0.11$\pm$0.02 &   2.92  \\

\multicolumn{6}{c}{[Fe/H]$=-$0.10$\pm$0.05 dex}           \\
  Stock 2 &   0.27$\pm$0.01 &   0.23$\pm$0.01 &   0.06$\pm$0.04 &  -0.09$\pm$0.04 &   3.60  \\
 NGC 6940 &   0.25$\pm$0.04 &   0.22$\pm$0.04 &   0.05$\pm$0.03 &  -0.05$\pm$0.03 &   2.33  \\
 NGC 2539 &   0.27$\pm$0.02 &   0.23$\pm$0.02 &   0.00$\pm$0.01 &  -0.07$\pm$0.01 &   2.89  \\
 NGC 7762 &   0.11$\pm$0.02 &   0.08$\pm$0.02 &   0.05$\pm$0.03 &  -0.08$\pm$0.03 &   1.50  \\
 NGC 2482 &   0.30$\pm$0.02 &   0.26$\pm$0.02 &   0.07$\pm$0.02 &  -0.03$\pm$0.02 &   2.82  \\
 NGC 2360 &   0.20$\pm$0.01 &   0.16$\pm$0.01 &   0.09$\pm$0.02 &   0.03$\pm$0.02 &   2.53  \\
 NGC 7209 &   0.27$\pm$0.03 &   0.23$\pm$0.03 &   0.06$\pm$0.02 &  -0.06$\pm$0.02 &   2.79  \\
 NGC 2682 &   0.25$\pm$0.02 &   0.22$\pm$0.02 &   0.09$\pm$0.01 &  -0.01$\pm$0.01 &   1.20  \\
 NGC 2251 &   0.33$\pm$0.03 &   0.29$\pm$0.03 &   0.00$\pm$0.03 &  -0.11$\pm$0.03 &   3.21  \\
 NGC 1912 &   0.33$\pm$0.05 &   0.28$\pm$0.05 &   0.06$\pm$0.02 &  -0.03$\pm$0.02 &   3.13  \\
 NGC 2527 &   0.32$\pm$0.03 &   0.28$\pm$0.03 &   0.05$\pm$0.02 &  -0.02$\pm$0.02 &   2.71  \\
 NGC 1662 &   0.22$\pm$0.01 &   0.17$\pm$0.01 &  -0.03$\pm$0.03 &  -0.09$\pm$0.03 &   2.77  \\
 NGC 2548 &   0.26$\pm$0.03 &   0.22$\pm$0.03 &   0.03$\pm$0.02 &  -0.03$\pm$0.02 &   2.83  \\
 NGC 2168 &   0.22$\pm$0.02 &   0.21$\pm$0.02 &   0.03$\pm$0.03 &  -0.11$\pm$0.03 &   4.60  \\
 NGC 1664 &   0.27$\pm$0.02 &   0.23$\pm$0.02 &  -0.01$\pm$0.02 &  -0.09$\pm$0.02 &   3.13  \\
 NGC 1817 &   0.16$\pm$0.02 &   0.12$\pm$0.02 &   0.11$\pm$0.02 &   0.04$\pm$0.02 &   2.80  \\
 NGC 2447 &   0.12$\pm$0.01 &   0.09$\pm$0.01 &  -0.14$\pm$0.03 &  -0.23$\pm$0.03 &   2.85  \\
 NGC 1342 &   0.28$\pm$0.02 &   0.24$\pm$0.02 &  -0.05$\pm$0.02 &  -0.11$\pm$0.02 &   2.71  \\
 NGC 1647 &   0.37$\pm$0.05 &   0.36$\pm$0.05 &   0.13$\pm$0.05 &  -0.02$\pm$0.05 &   4.09  \\

\multicolumn{6}{c}{[Fe/H]$=-$0.20$\pm$0.05 dex}           \\
 NGC 2437 &   0.32$\pm$0.03 &   0.28$\pm$0.03 &  -0.01$\pm$0.02 &  -0.13$\pm$0.02 &   3.29  \\
 NGC 2354 &   0.12$\pm$0.05 &   0.09$\pm$0.05 &  -0.11$\pm$0.03 &  -0.18$\pm$0.03 &   4.24  \\
 NGC 2335 &   0.24$\pm$0.02 &   0.20$\pm$0.02 &  -0.02$\pm$0.02 &  -0.08$\pm$0.02 &   3.90  \\
 NGC 2287 &   0.24$\pm$0.03 &   0.23$\pm$0.03 &   0.05$\pm$0.02 &  -0.08$\pm$0.02 &   3.29  \\
 NGC 2506 &   0.21$\pm$0.04 &   0.18$\pm$0.04 &   0.17$\pm$0.01 &   0.09$\pm$0.01 &   2.02  \\
 NGC 2345 &   0.18$\pm$0.04 &   0.18$\pm$0.04 &   0.03$\pm$0.02 &  -0.10$\pm$0.02 &   2.50  \\

\hline
\end{tabular}
\end{table*}


\begin{thebibliography}{99}

 \bibitem{} Alexeeva S. A., Pakhomov Yu. V., Mashonkina L. I., 2014, AstL, 40, 406
 \bibitem{} Alonso A., Arribas S., Mart\'{i}nez-Roger C., 1999, A\&AS, 140, 261
 \bibitem{} Armillotta L., Krumholz M. R., Fujimoto Y., 2018, MNRAS, 481, 5000
 \bibitem{} Arnould M., Goriely S., Jorissen A., 1999, A\&A, 347, 572
 \bibitem{} Asplund M., Grevesse N., Sauval A. J., Scott P., 2009, ARA\&A, 47, 481
 \bibitem{} Asplund M., Grevesse N., Sauval A. J., 2005, ASPC, 336, 25
 \bibitem{} Bagdonas V., Drazdauskas A., Tautvai{\v s}ien{\. e} G., Smiljanic R., Chorniy Y., 2018, A\&A, 615, 165
 \bibitem{} Barrado y Navascu\'{e}s D., Deliyannis C. P., Stauffer J. R., 2001, ApJ, 549, 452
 \bibitem{} Battistini C., Bensby T., 2016, A\&A, 586, 49
 \bibitem{} Bertelli M. C., Pasquali A., Richer J., Michaud G., Salaris M., et al. 2018, MNRAS, 478, 425
 \bibitem{} Bisterzo S., Travaglio C., Gallino R., Wiescher M., K{\: a}ppeler F., 2014, ApJ, 787, 10
 \bibitem{} Blanco-Cuaresma S., Soubiran C., Heiter U., Asplund, M., et al. 2015, A\&A, 577, 47
 \bibitem{} Bragaglia A., Fu X., Mucciarelli A., Andreuzzi G., Donati P., 2018, A\&A, 619, 176
 \bibitem{} Cantat-Gaudin T., Jordi C., Vallenari A., Bragaglia A., et al. 2018, A\&A, 618, 93
 \bibitem{} Carraro G., Semenko E. A., Villanova S., 2016, AJ, 152, 224
 \bibitem{} Carretta E., Bragaglia A., Gratton R. G., Tosi M., 2005, A\&A, 441, 131
 \bibitem{} Casamiquela L., Carrera R., Blanco-Cuaresma S., Jordi C., et al. 2017, MNRAS, 470, 4363
 \bibitem{} Casamiquela L., Carrera R., Jordi C., Balaguer-N\'{u}\v{n}ez L., et al.  2016, MNRAS, 458, 3150
 \bibitem{} Castelli F., Kurucz R. L., 2003, IAU Symposium 210, Modelling of Stellar Atmospheres, Uppsala, Sweden, eds. N.E. Piskunov, W.W. Weiss, and D. F. Gray, 2003, ASP-S210
 \bibitem{} Charbonnel C., Lagarde N., 2010, A\&A, 522, 10
 \bibitem{} Cunha K., Hubeny I., Lanz T., 2006, ApJ, 647, 143
 \bibitem{} Cutri R. M., Skrutskie M. F., van Dyk S., et al. 2003, The IRSA 2MASS All-Sky Point Source Catalog (Pasadena, CA: IPAC/California Institute of Technology)
 \bibitem{} Delgado Mena E., Tsantaki M., Adibekyan V. Zh., Sousa S. G., Santos N. C., Gonz\'{a}lez Hern\'{a}ndez J. I., Israelian G., 2017, A\&A, 606, 94
 \bibitem{} Drazdauskas A., Tautvai\v{s}ien\.{e} G., Smiljanic R., Bagdonas V., Chorniy Y., 2016, MNRAS, 462, 794
 \bibitem{} Ferraro F., Tak{\' a}cs M. P., Piatt{\. i} D., Cavanna F., et al. 2018, PhRvL, 121, 2701
 \bibitem{} Foster D. C., Theissen A., Butler C. J., Rolleston W. R. J., Byrne P. B., Hawley S. L., 2000, A\&AS, 143, 409
 \bibitem{} Freeman K., Bland-Hawthorn J., 2002, ARA\&A, 40, 4875
 \bibitem{} Gaia Collaboration et al. 2018, A\&A, 616, 1
 \bibitem{} Gao X., Lind K., Amarsi A. M., Buder S., et al. 2018, MNRAS, 481, 2666
 \bibitem{} Geller A.~M., Latham D.~W, Mathieu R.~D., 2015, AJ, 150, 97
 \bibitem{} Geller A.~M., Mathieu R.~D., Braden E.~K., et al. 2010, AJ, 139, 1383
 \bibitem{} Girard T.~M., Grundy W.~M., Lopez C.~E., van Altena W.~F., 1989, AJ, 98, 227
 \bibitem{} Gratton R.G., Carretta E., Eriksson K., Gustafsson B., 1999, A\&A, 350, 955
 \bibitem{} Hamdani S., North P., Mowlavi N., Raboud D., Mermilliod J.-C., 2000, A\&A, 360, 509
 \bibitem{} Jordi C., Gebran M., Carrasco J. M., de Bruijne J., et al. 2010, A\&A, 523, 48
 \bibitem{} Karakas A. I., Lattanzio J. C., 2014, PASA, 31, 30 
 \bibitem{} Kharchenko N. V., et al. 2001, Kinematics and Physics of Celestial Bodies, 17, 409
 \bibitem{} Kharchenko N. V., Piskunov A. E., R{\"o}ser S., Schilbach E., Scholz R.-D., 2005, A\&A, 438, 1163
 \bibitem{} Kurucz R. L., Furenlid I., Brault J., \& Testerman L., 1984, Solar Flux Atlas from 296 to 1300 nm, ed. R. L. Kurucz, I. Furenlid, J. Brault, \& L. Testerman (Sunspot, NM: National Solar Observatory)  
 \bibitem{} Krzeminski W., Serkowski K., 1967, ApJ, 147, 988
 \bibitem{} Lagarde N., Decressin T., Charbonnel C., Eggenberger P., Ekström S., Palacios A., 2012, A\&A, 543, 108
 \bibitem{} Lambert D. L., Reddy A. B. S., 2016, ApJ, 831, 202
 \bibitem{} Lind K., Asplund M., Barklem P. S., Belyaev A. K., 2011, A\&A, 528, 103
 \bibitem{} Luck R.~E. 2015, AJ, 150, 88
 \bibitem{} Maciejewski G., Boeva S., Georgiev Ts., Mihov B., Ovcharov E., Valcheva A., Niedzielski A., 2008, BaltA, 17, 51
 \bibitem{} Magrini L., Spina L., Randich S., Friel E., Kordopatis G., et al. 2018, A\&A, 617, 106
 \bibitem{} Marigo P., Girardi L., Bressan A., Groenewegen M. A. T., Silva L., Granato G. L., 2008, A\&A, 482, 883
 \bibitem{} Mathieu R. D. 2000, in ASPC, Vol. 198, ``Stellar Clusters and Associations: Convection, Rotation, and Dynamos'', ed. R. Pallavicini, G. Micela, \& S. Sciortino, 517
 \bibitem{} McNamara B. J. Sekiguchi K., 1986a, AJ, 91, 557
 \bibitem{} Mermilliod J.-C., Mayor M., Udry S., 2009, A\&A, 498, 949
 \bibitem{} Mermilliod J.-C., Mayor M., Udry S., 2008, A\&A, 485, 303
 \bibitem{} Mishenina T.~V., Bienaym{\'e} O., Gorbaneva T.~I., et~al. 2006, A\&A, 456, 1109
 \bibitem{} Mishenina T.~V., Gorbaneva T.~I., Bienaym{\'e} O., et~al. 2007, Astronomy Reports, 51, 382
 \bibitem{} Mishenina T., Kovtyukh V., Soubiran C., Adibekyan V. Zh, 2016, MNRAS, 462, 1563
 \bibitem{} Mowlavi N., 1999, A\&A, 350, 73
 \bibitem{} Nordlander T., Lind K., 2017, A\&A, 607, A75
 \bibitem{} Prosser C. F., Stauffer J. R., Caillault J.-P., Balachandran S., Stern R. A.,  Randich S., 1995, AJ, 110, 1229
 \bibitem{} Reddy A. B. S., Giridhar S., Lambert D. L., 2015, MNRAS, 450, 4301
 \bibitem{} Reddy A. B. S., Giridhar S., Lambert D. L., 2013, MNRAS, 431, 3338 
 \bibitem{} Reddy A. B. S., Giridhar S., Lambert D. L., 2012, MNRAS, 419, 1350
 \bibitem{} Reddy A. B. S., Lambert D. L., 2017, ApJ, 845, 151
 \bibitem{} Reddy A. B. S., Lambert D. L., 2015, MNRAS, 454, 1976
 \bibitem{} Reddy A. B. S., Lambert D. L., Giridhar S., 2016, MNRAS, 463, 4366
 \bibitem{} Roeser S., Demleitner M., Schilbach E., 2010, AJ, 139, 2440
 \bibitem{} Skrutskie M. F., Cutri R. M., Stiening R., et al.  2006, AJ, 131, 1163
 \bibitem{} Smiljanic R., 2012, MNRAS, 422, 1562
 \bibitem{} Smiljanic R., Donati P., Bragaglia A., Lemasle B., Romano D., 2018, A\&A, 616, 112
 \bibitem{} Smiljanic R., Romano D., Bragaglia A., Donati P., 2016, A\&A, 589, 115
 \bibitem{} Sneden C., 1973, PhD Thesis, Univ. of Texas, Austin
 \bibitem{} Sneden C., Cowan J., Gallino R., 2008, ARA\&A, 46, 241
 \bibitem{} Soubiran C., Cantat-Gaudin T., Romero-G\'{o}mez M., Casamiquela L., et al. 2018, A\&A, 619, 155
 \bibitem{} Souto D., Cunha K., Smith V. V., Allende Prieto C., et al. 2018, ApJ, 857, 14
 \bibitem{} Spagna A., Cossu F., Lattanzi M.~G., Massone G., 2009, MmSAI, 80, 129
 \bibitem{} Sung H., Bessell M. S., 1999, MNRAS, 306, 361
 \bibitem{} Tull R.G., MacQueen P.J., Sneden C., Lambert D.L., 1995, PASP, 107, 251
 \bibitem{} Ventura P., Di~Criscienzo M., Carini R., D'Antona F., 2013, MNRAS, 431, 3642
 \bibitem{} Villanova S., Carraro G., Saviane I., 2009, A\&A, 504, 845


\end{thebibliography}
\end{document}